\documentclass[acmsmall]{acmart}

\usepackage{listings}
\usepackage[capitalise,nameinlink]{cleveref}
\usepackage[export]{adjustbox}
\usepackage[many]{tcolorbox}

\usepackage{enumitem}
\usepackage{pifont}
\usepackage{subcaption}
\usepackage{xspace}
\usepackage{multirow}

\newcommand{\cmark}{\ding{51}} 
\newcommand{\C}[1]{\texttt{\small#1}}

\newcommand\wasms[1]{{\small\K{#1}}\xspace}
\newcommand{\codens}[1]{\texttt{\small #1}}
\newcommand{\code}[1]{{\small\texttt{#1}}\xspace}
\newcommand{\web}{web\xspace}
\newcommand\XX{$\times$\xspace}
\newcommand{\name}[1]{\textsc{#1}}

\definecolor{RoyalBlue}{HTML}{0071BC}
\definecolor{ForestGreen}{HTML}{009B55}
\definecolor{LinkColor}{rgb}{0.55,0.0,0.3}
\definecolor{CiteColor}{rgb}{0.55,0.0,0.3}
\definecolor{HighlightColor}{rgb}{0.0,0.0,0.0}
\definecolor{Gray}{gray}{0.9}
\definecolor{grey}{rgb}{0.5,0.5,0.5}
\definecolor{red}{rgb}{1,0,0}
\definecolor{darkgreen}{rgb}{0.0,0.7,0.0}
\definecolor{backcolor}{rgb}{0.95,0.95,0.95}

\definecolor{keywordcolor}{RGB}{0,0,255}
\definecolor{commentcolor}{RGB}{0,128,0}
\definecolor{stringcolor}{RGB}{163,21,21}

\lstdefinelanguage{WebAssembly}{
  keywords={i32, i64, f32, f64, const, get, store, add, tee, global, local, load, store8, set, module, func, call, drop, memory, export, import, param, result},
  keywordstyle=\color{keywordcolor}\bfseries,
  comment=[l]{;},
  commentstyle=\color{commentcolor},
  string=[b]",
  stringstyle=\color{stringcolor},
  sensitive=true,
  morecomment=[l]{;;},
  morestring=[b]',
  morestring=[b]"
}

\lstdefinelanguage{JavaScript}{
  keywords={typeof, new, true, false, catch, function, return, null, catch, switch, var, if, in, while, do, else, case, break, const, let, await},
  keywordstyle=\color{blue}\bfseries,
  ndkeywords={class, export, boolean, throw, implements, import, this},
  ndkeywordstyle=\color{darkgray}\bfseries,
  identifierstyle=\color{black},
  sensitive=false,
  comment=[l]{//},
  morecomment=[s]{/*}{*/},
  commentstyle=\color{purple}\ttfamily,
  stringstyle=\color{red}\ttfamily,
  morestring=[b]',
  morestring=[b]"
}

\lstdefinelanguage{Definition}{
  keywords={}, 
  keywordstyle=\color{green}\bfseries,
  ndkeywords={class, export, boolean, throw, implements, import, this},
  ndkeywordstyle=\color{darkgray}\bfseries,
  identifierstyle=\color{black},
  sensitive=false,
  comment=[l]{//},
  morecomment=[s]{/*}{*/},
  commentstyle=\color{purple}\ttfamily,
  stringstyle=\color{red}\ttfamily,
  morestring=[b]',
  morestring=[b]"
}

\hypersetup{%
  linktocpage=true, pdfstartview=FitV,
  breaklinks=true, pageanchor=true, pdfpagemode=UseOutlines,
  plainpages=false, bookmarksnumbered, bookmarksopen=true, bookmarksopenlevel=3,
  hypertexnames=true, pdfhighlight=/O,
  colorlinks=true,linkcolor=LinkColor,citecolor=CiteColor,
  urlcolor=LinkColor
}

\crefformat{section}{#2\S{}#1#3}

\tcolorboxenvironment{lstlisting}{
  spartan,
  frame empty,
  boxsep=0mm,
  left=0mm,right=0mm,top=-1mm,bottom=-1mm,
  colback=backcolor,
}
\lstdefinestyle{mystyle}{
  keywordstyle=\color{RoyalBlue},
  commentstyle=\color{ForestGreen},
  basicstyle=\linespread{0.9}\ttfamily\footnotesize,
  breakatwhitespace=false,
  breaklines=true,
  captionpos=b,
  keepspaces=true,
  numbers=left,
  numbersep=10pt,
  numberstyle=\scriptsize\ttfamily\color{grey},
  xleftmargin=18pt,
  showspaces=false,
  showstringspaces=false,
  escapechar=$,
}
\lstdefinelanguage{Rust}{
  keywords={as,break,const,continue,crate,else,enum,extern,false,fn,for,if,
  impl,in,let,loop,match,mod,move,mut,pub,ref,return,self,Self,static,struct,
  super,trait,true,type,unsafe,use,where,while,async,await,dyn,abstract,become,
  box,do,final,macro,override,priv,typeof,unsized,virtual,yield,try},
  morecomment=[l]{//},
  morecomment=[s]{/*}{*/},
}

\setcopyright{rightsretained}
\acmDOI{10.1145/3689787}
\acmYear{2024}
\acmJournal{PACMPL}
\acmVolume{8}
\acmNumber{OOPSLA2}
\acmArticle{347}
\acmMonth{10}
\acmSubmissionID{oopslab24main-p738-p}
\received{2024-04-06}
\received[accepted]{2024-08-18}
\begin{document}

\title{Wasm-R3: Record-Reduce-Replay for Realistic and Standalone WebAssembly Benchmarks}

\author{Doehyun Baek}
\authornote{Both authors contributed equally to this research.}
\email{doehyun.baek@kaist.ac.kr}
\orcid{0009-0004-0117-1060}
\affiliation{
  \institution{KAIST}
  \city{Daejeon}
  \country{South Korea}
}

\author{Jakob Getz}
\authornotemark[1]
\email{jakob@getz.de}
\orcid{0009-0009-7656-2329}
\affiliation{
  \institution{University of Stuttgart}
  \city{Stuttgart}
  \country{Germany}
  }

\author{Yusung Sim}
\email{yusungsim@kaist.ac.kr}
\orcid{0000-0003-3641-593X}
\affiliation{
  \institution{KAIST}
  \city{Daejeon}
  \country{South Korea}
}

\author{Daniel Lehmann}
\email{mail@dlehmann.eu}
\orcid{0000-0002-4037-5152}
\affiliation{
 \institution{Google Germany GmbH}
 \city{Munich}
 \country{Germany}
 }

\author{Ben L. Titzer}
\orcid{0000-0002-9690-2089}
\email{btitzer@andrew.cmu.edu}
\affiliation{
  \institution{Carnegie Mellon University}
  \city{Pittsburgh}
  \country{USA}
  }

\author{Sukyoung Ryu}
\orcid{0000-0002-0019-9772}
\email{sryu.cs@kaist.ac.kr}
\affiliation{
  \institution{KAIST}
  \city{Daejeon}
  \country{South Korea}
  }

\author{Michael Pradel}
\orcid{0000-0003-1623-498X}
\email{michael@binaervarianz.de}
\affiliation{%
  \institution{University of Stuttgart}
  \city{Stuttgart}
  \country{Germany}
  }

\begin{abstract}
  WebAssembly (Wasm for short) brings a new, powerful capability to the \web as well as Edge, IoT, and embedded systems.
  Wasm is a portable, compact binary code format with high performance and robust sandboxing properties.
  As Wasm applications grow in size and importance, the complex performance characteristics of diverse Wasm engines demand robust, representative benchmarks for proper tuning.
  Stopgap benchmark suites, such as PolyBenchC and libsodium, continue to be used in the literature, though they are known to be unrepresentative.
  Porting of more complex suites remains difficult because Wasm lacks many system APIs and extracting real-world Wasm benchmarks from the \web is difficult due to complex host interactions.
  To address this challenge, we introduce \emph{Wasm-R3}, the first record and replay technique for Wasm.
  Wasm-R3 transparently injects instrumentation into Wasm modules to \emph{record} an execution trace from inside the module, then \emph{reduces} the execution trace via several optimizations, and finally produces a \emph{replay} module that is executable standalone without any host environment---on any engine.
  The benchmarks created by our approach are (i) realistic, because the approach records real-world \web applications, (ii) faithful to the original execution, because the replay benchmark includes the unmodified original code, only adding emulation of host interactions, and (iii) standalone, because the replay benchmarks run on any engine.
  Applying Wasm-R3 to \web-based Wasm applications in the wild demonstrates the correctness of our approach as well as the effectiveness of our optimizations, which reduce the recorded traces by 99.53\% and the size of the replay benchmark by 9.98\%.
  We release the resulting benchmark suite of 27 applications, called \emph{Wasm-R3-Bench}, to the community, to inspire a new generation of realistic and standalone Wasm benchmarks.  
\end{abstract}

\ccsdesc[500]{Software and its engineering~Software maintenance tools}

\keywords{WebAssembly, Benchmarking, record and replay}
\maketitle

\section{Introduction}
\label{sec:introduction}
WebAssembly (Wasm) is a portable, low-level code format designed for compact representation and efficient sandboxed execution.
It is primarily used as a compilation target for various source languages, including C/C++, Rust, and Kotlin, enabling new classes of software to be run in the browser.
Its low-level instructions closely map to hardware instructions, achieving near-native performance with straightforward compilation techniques.
Wasm augments the \web platform, promising to speed up specific components of broader applications~\cite{WebAssemblyIntroduction}, as Wasm code often runs faster than JavaScript for numeric and memory-intensive tasks.

The complexity and diversity of Wasm engines demand robust, representative benchmarks for proper tuning.
On the \web, Wasm code is loaded dynamically from URLs or can be dynamically generated.
Thus, code processing time, interpretation overhead, and JIT compilation to native code contribute to the overall application run time.
To deliver fast startup time and high peak performance, all of today's \web browsers employ multi-tier Wasm engines.
For example, V8 and SpiderMonkey, the engines used in Chrome and Firefox respectively, use two compiler tiers~\cite{v8liftoff, TurboFan}, while JavaScriptCore in Safari uses an interpreter and two compiler tiers.
Non-\web Wasm engines often also use multiple tiers, such as the Wizard Research Engine~\cite{wizard} which employs a new in-place interpreter design and a baseline compiler~\cite{CGO2024}.
Multi-tier engines have complex performance characteristics and their tiering heuristics need to be tuned on realistic applications to ensure both startup speed and peak performance are maximized.
Tuning these systems requires large, complex workloads that are representative of real-world applications.
In the past, unrepresentative benchmarks, such as SunSpider for JavaScript~\cite{Richards2010AnAO}, have misdirected engineering effort. In one instance, bad benchmarks led engineers into believing that their performance optimization resulted in a 13× improvement, when a representative benchmark showed only a 3× improvement \cite{jsbench}.

Unfortunately, creating a sufficiently large set of representative Wasm workloads is challenging.
One possible approach might be to port existing native applications to Wasm.
However, the lack of standardized system APIs has made porting or recompiling large native applications difficult, as efforts like WASI~\cite{WASI} still lack basic facilities, such as signals, sockets, permissions, shared memory, and device APIs.
Moreover, beyond the three major \web engines, there are now several non-\web production Wasm engines, which support disparate APIs or only subsets of WASI, making it difficult to have complex benchmarks with non-trivial system interaction.
This deficiency in Wasm benchmarks compels researchers to either write their own benchmarks or use commonly used standard benchmark suites with few dependencies, such as PolyBenchC~\cite{PolyBenchC} and libsodium~\cite{LibSodiumBench}, which have already been shown to be unrepresentative of real-world applications~\cite{notSoFast}.
Another approach might be to extract benchmarks from real-world \web applications.
However, Wasm-based \web applications consist not only of Wasm modules, but also of a host environment that runs JavaScript code, interacts with the network, and interacts with the user.
These properties make the \web host environment difficult to emulate, which poses a problem for creating benchmark suites from representative \web applications.

\begin{table}[t]
  \centering
\small
\caption[Real world websites evaluation set.]{\label{t:source-level}
A comparison of record and replay frameworks.}
\vspace*{-.5em}
  \begin{tabular}{l|cccc}
  \hline
  \textbf{} & \textbf{PinPlay~\cite{PinPlay}} & \textbf{JSBench~\cite{jsbench}} & \textbf{Jalangi~\cite{jalangi}} & \textbf{Wasm-R3 (this work)}\\
  \hline
  Cross-architecture replay &  & \cmark & \cmark & \cmark \\
  Cross-language replay & \cmark &  &  & \cmark \\
  Accurate replay & \cmark &  & \cmark & \cmark \\
  Code format & native binaries & JavaScript & JavaScript & WebAssembly  \\
  \hline
  \end{tabular}
\vspace*{-1em}
\end{table}

Outside of the Wasm context, creating good benchmarks is a long-standing challenge in many areas of systems.
Efforts span virtual machines, operating systems, architecture, vision, and machine learning~\cite{dacapo,SPEC,MiBench,MLBench,Speedometer3}.
Key considerations are the size and runtime of benchmarks, ease of compiling and running them, licensing of the underlying source or binaries, diversity of the suite, representativeness of chosen benchmarks, and standardized measurement and reporting methodologies.
For nascent and developing domains, writing new benchmarks makes sense, but for established domains with real-world usage, benchmarks should reflect actual applications to direct tuning efforts to benefit real-world usage.
One such benchmark for Wasm is PSPDFKit~\cite{pspdfkit}, which measures the runtime of different actions of a PDF library.
Yet, despite being well-crafted, this benchmark still depends on \web APIs and requires significant effort to disentangle it to run on other engines.
Moreover, creating and maintaining benchmarks like this requires significant manual engineering effort, so few examples exist so far.


If creating and curating benchmarks requires so much manual effort, why not automate the process?
A promising approach is to automatically record and replay executions of real-world applications.
Indeed, record and replay techniques for several systems and languages have been proposed, as shown in Table~\ref{t:source-level}.
However, each of these techniques lacks in a different dimension.
PinPlay~\cite{PinPlay} supports recording of execution across multiple architectures, but cannot replay across different architectures.
Language-specific efforts like JSBench \cite{jsbench} and Jalangi \cite{jalangi} are portable across CPUs and OSes, but only serve one language.
Recording JavaScript is challenging; JSBench cannot record all memory loads, which means the execution at replay might diverge from the original execution \cite{jalangi}.
Moreover, these techniques are not directly applicable to Wasm, and to the best of our knowledge, there currently is no record and replay technique for Wasm.



To address the lack of realistic benchmarks for Wasm, we present \emph{Wasm-R3}, the first record and replay technique for Wasm that enables the creation of benchmarks from executions of real-world applications.
Our key insight is that the design of Wasm modules enforces a clear separation between imported host functionality and the state and behavior inside a Wasm module, and that this is a natural boundary for encapsulating a benchmark. 
Our Wasm-\underline{R3} approach consists of three phases: \underline{r}ecord, \underline{r}educe, and \underline{r}eplay.
To \emph{record} an execution, the approach transparently injects instrumentation into Wasm modules and records all interactions with the environment.
Because na\"{i}vely recording all interactions would result in an impractically large trace, Wasm-R3 \emph{reduces} the trace via several optimizations.
Finally, Wasm-R3 produces a \emph{replay} benchmark that contains the unmodified code of the original Wasm module, but factors out the host environment and replaces it with a replay mechanism included directly in the replay benchmark.

Recording and replaying at the module boundary is akin to techniques for replaying native binaries at the system-call layer.
Yet unlike native binary replay techniques~\cite{PinPlay,valgrind,dynamorio}, which often use memory-checkpointing techniques, the diverse host environments and engine offerings for Wasm demand a more general technique that works with an {\it uncooperative} host environment.
The term ``uncooperative'' here means that the host environment does not provide any support for record and replay.
Instead, Wasm-R3 works without any modifications of the host environment or the underlying Wasm engine, but instruments a Wasm module so that it records its own trace.

The benchmarks created with Wasm-R3 are portable, i.e., they work wherever Wasm runs.
Since Wasm is gaining adoption across a broad range of contexts, such as Cloud~\cite{Mntrey2022WebAssemblyAA}, Edge~\cite{Sledge}, and IoT~\cite{WAIT}, Wasm-R3 must work across architectures, operating systems, runtimes, and host environments.
While some record and replay techniques rely on support by the hardware~\cite{FDR}, the operating system~\cite{revirt}, or the language runtime system~\cite{RANDR}, the vast diversity of Wasm means that no one of these techniques can apply to all Wasm environments.
Instead, Wasm-R3 produces self-contained, standalone Wasm modules that replay their execution faithfully not only on the engine used to create the benchmark, but on any engine.
Moreover, the produced modules include the unmodified functions from the original Wasm module, only adding replay functions.
Since the technique is primarily additive, the performance characteristics of the original functions are similar.
We show in experiments that nearly all benchmarks produced with Wasm-R3 spend the majority of their execution time in the original functions, not in replay code.


Our evaluation applies Wasm-R3 to real-world \web-based Wasm applications, demonstrating that the approach is effective in creating 27 realistic and standalone benchmarks.
We show that our optimizations effectively reduce the size of the recorded trace (by 99.53\%, on average) and the size of the replay benchmark (by 9.98\%, on average).
The generated replay code accounts only for 0.20\% of the total execution time, and hence, the extracted benchmarks accurately represent the original application.
We release the benchmark suite created by Wasm-R3 during our evaluation, called \emph{Wasm-R3-Bench}, to the community, and envision them to serve as a new standard for realistic and standalone Wasm benchmarks.

In summary, this paper contributes the following:
\begin{itemize}
    \item We introduce the first record and replay technique for Wasm. It does not require support from or modification of the Wasm host environment, hardware, operating system, language runtime, or source compiler.
    \item We demonstrate the technique via a system which records execution traces of \web applications in any browser and produces replay benchmarks that execute without any host environment---on any engine.
    \item We present optimization techniques that reduce trace size and improve replay performance.
    For several applications, these optimizations are vital to making our approach feasible at all, avoiding out-of-memory errors and excessive slowdowns that make benchmarks unrepresentative of the original application's performance.
    \item We demonstrate that Wasm-R3 is effective in real-world scenarios by using the approach to create benchmarks from 27 real-world \web applications.
    \item We make Wasm-R3, associated tools, and the created benchmarks available as open source \url{https://github.com/sola-st/wasm-r3}.
\end{itemize}

\section{Background}
\label{sec:background}


\newcommand\X[1]{\mathit{#1}}
\newcommand\K[1]{\textsf{\textbf{#1}}}

\begin{figure}[t]
\def\baselinestretch{1.05}

\footnotesize

$$
\begin{array}{@{}rrlrrl@{}}
\X{module} &::=& \X{function}^\ast~\X{global}^\ast~\X{start}^?~\X{table}^?~\X{memory}^?
 & \X{import}, \X{export} &::=& \C{\footnotesize "name"} \\
\X{function} &::=& \X{type_{func}}~(\X{import}~|~\X{code})~\X{export}^\ast
 & \X{code} &::=& \K{(local}~\X{type_{val}})^\ast~\X{instr}^\ast \\
\X{global} &::=& \X{type_{val}}~(\X{import}~|~\X{init})~\X{export}^\ast
 & \X{init} &::=& \X{instr}^\ast \\
\X{start} &::=& \X{idx_{func}} & 
\X{type_{val}} &::=& \K{i32}~\mid~\K{i64}~\mid~\K{f32}~\mid~\K{f64} \\
\X{table} &::=& \X{import}^?~\X{idx_{func}}^\ast~\X{export}^\ast &
\X{type_{func}} &::=& \X{type_{val}}^\ast \rightarrow \X{type_{val}}^\ast \\
\X{memory} &::=& \X{import}^?~\C{\footnotesize byte}^\ast~\X{export}^\ast &
\X{idx_{func\,|\,global\,|\,local}} &\in & \mathbb{N} \\
\X{instr} &::=& \lefteqn{
\X{type_{val}}\K{.const}~\X{value}~\mid~
\X{type_{val}}\K{.load}~\mid~\X{type_{val}}\K{.store}~\mid~\K{memory.grow}} \\ 
&\mid& \lefteqn{\K{call}~\X{idx_{func}}~\mid~\K{call\_indirect}~\X{type_{func}}~\mid~\K{return}~\mid~\cdots} \\
\end{array}
$$
\vspace*{-0.7em}
\caption{Excerpt from the abstract syntax of a simplified form of Wasm~\cite{Wasabi}.}\label{t:mini-wasm}
\vspace*{-1em}
\end{figure}

In this section, we provide necessary information to describe Wasm-R3.
Figure~\ref{t:mini-wasm} shows an excerpt of a simplified abstract syntax of Wasm.
%
A Wasm \emph{module} denotes a single binary file and
consists of functions, global variables, an optional start function,
and one table and memory.
A \emph{function} takes parameters, declares local variables, executes body instructions, and returns a sequence of results.
A \emph{global variable} stores a single value and can be accessed from all functions and can be either mutable or immutable.
A \emph{start function} is automatically executed when the module is loaded.
A \emph{table} maps function indices to opaque references to either extern (host) objects or Wasm functions.
Tables can be used for indirect function calls via the \wasms{call\_indirect} instruction.
A \emph{memory} represents a contiguous, byte-addressable, page-sized mutable array of memory.
All of these entities can either be imported from a host execution environment, specifying a module and name pair, or exported under one or more names, allowing them to be accessed externally.
Apart from these entities, modules can include initialization data for tables and memories.

One concept that plays a key role in Wasm-R3 is embedding of Wasm modules into a host environment.
Host environments load Wasm modules, resolve imports and exports between modules, and provide host functions as imports to Wasm functions.
Host functions can access state outside modules and perform I/O.
While the web was the main motivation for Wasm initially, it was designed to be embedded in multiple environments \cite{WebAssemblyIntroduction}.
Thus, although the web and JavaScript embedding was the primary one at launch, WASI~\cite{WASI} has emerged as a set of standard system APIs in non-web use cases.
In principle, unlimited embedders are possible due to the environment-agnostic design of Wasm.
After an embedder loads, verifies, and processes Wasm code, it provides bindings to a module's imports and creates the module's storage.
The result is an \emph{instance}, a runtime representation that contains the state of the module.
A Wasm instance interacts with host environments by calling imported host functions and being called by exported Wasm functions.

Consideration of host functions introduces another important aspect of Wasm for Wasm-R3: nondeterminism.
Since its inception, one of Wasm's explicit goals \cite{WebAssemblyIntroduction} has been to provide deterministic semantics across different hardware.
However, there are three exceptions: NaN payloads, resource exhaustion, and host functions.
Some Wasm instructions output non-deterministic NaN bit patterns in the presence of non-canonical input NaNs, as hardware may behave differently and canonicalizing all NaNs is deemed too expensive.
Resources like memory obviously vary from host to host and computer to computer, so deep recursion or \wasms{memory.grow} might fail at different points, and of course a host function can perform I/O or even arbitrary updates to a Wasm instance's exported state.
In this work, we focus on nondeterminism arising from the interaction with host functions.

\section{Approach}
\label{sec:approach}

\subsection{Overview}
\label{s:overview}

\begin{figure}[t]
  \centering
  \includegraphics[width=\linewidth,trim=0 20pt 0 0]{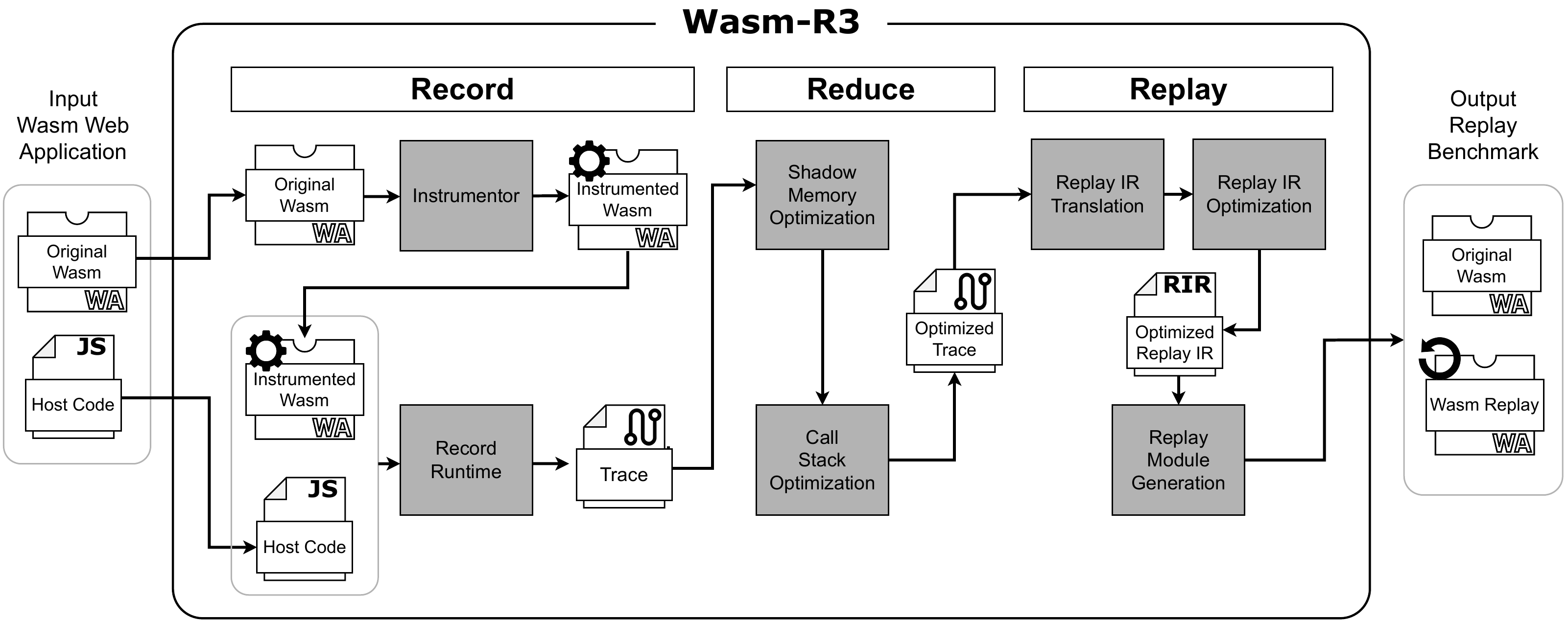} 
  \caption{Overview of the Wasm-R3 main phases and components.}
\label{fig:overview}
\end{figure}

Figure~\ref{fig:overview} gives an overview of Wasm-R3.
Given a Wasm-based web application, Wasm-R3 executes the application, possibly with user input, and produces a benchmark that replays that execution without any user input.
Wasm-R3 consists of three phases: \emph{record}, \emph{reduce}, and \emph{replay}.

The record phase can be considered the frontend of Wasm-R3. 
It takes a Wasm web application as an input and produces a \emph{trace} as a result.
We assume that the application consists of Wasm modules and host code, e.g., in JavaScript.
As the first step, Wasm-R3 intercepts the Wasm modules before they are loaded into the application and instruments the modules with recording logic.
Our instrumentation tracks all function calls and returns as well as all loads and stores to mutable state.
Then, the instrumented Wasm web application is run within the \emph{record runtime}.
While the user interacts with the web application, each instrumented Wasm module records its own execution trace.
Note that if the input Wasm web application loads multiple Wasm modules, then the record phase produces one trace for each module. 

For complex applications, the traces can grow prohibitively large.
Thus, for efficiency, the reduce phase filters out unnecessary events from the output trace of the record phase.
Specifically, it applies the \emph{shadow memory optimization} and \emph{call stack optimization} on the trace.
By applying these optimizations, we only keep the parts of traces that are directly related to the nondeterminism that occured during execution.
Eventually, the reduce phase yields optimized traces as outputs.

The replay phase is the backend of Wasm-R3.
It takes the original, uninstrumented Wasm module and the corresponding optimized trace as inputs and produces a self-contained, executable \emph{replay benchmark}.
The replay phase does not modify the original module's functions.
Instead, it simply merges them with generated \emph{replay functions} to complete the executable replay benchmark.
It first translates the input trace to an intermediate representation, the \emph{replay IR}.
Then, it applies \emph{replay IR optimizations} to reduce the size of the IR and ultimately the generated code, ensuring that the resulting binaries satisfy the size restrictions commonly imposed by engines.
Finally, it generates the replay benchmark from the optimized replay IR.
As we discuss in Section~\ref{s:replay_outputs},
our replay binary generator supports three different output formats. 
This approach allows replay benchmarks to be executed in diverse environments, including web browsers and standalone Wasm runtimes.

The next sections dive into the details of the record phase (Section~\ref{s:record}), trace reduction techniques (Section~\ref{s:trace_reduce}), and the replay phase (Section~\ref{s:replay}).

\subsection{Record Phase}\label{s:record}
The record phase is responsible for recording the necessary information into a trace for reconstructing the benchmark that deterministically replays the Wasm web application's behavior.
As described earlier, non-determinism gets introduced to Wasm applications by functions imported from the host environment.
The record phase thus needs to capture information about function calls across the boundary between the host environment and the target Wasm module.
This includes side effects to the Wasm module's mutable state, e.g., the memory section of the Wasm module, caused by host functions.

The following describes the format of traces (Section~\ref{s:trace-structure}) and how these traces are recorded via instrumentation (Section~\ref{s:instrumentation}).

\subsubsection{Trace Structure}
\label{s:trace-structure}

\begin{figure}
    \begin{lstlisting}[language=Definition]
Trace      = Seq<Event>
Event      = FuncEntry | FuncReturn | Call | CallReturn | Load | Store
FuncEntry  = { funcidx: I32, params: Seq<ValType> }
FuncReturn = { funcidx: I32, values: Seq<ValType> }
Call       = { funcidx: I32 }
CallReturn = { funcidx: I32, results: Seq<ValType> }
Load       = { memidx: I32,
               address: I32,
               value: ValType | I8 | I16 }
Store      = { memidx: I32,
               address: I32,
               value: ValType | I8 | I16 }
ValType    = I32 | I64 | F32 | F64
\end{lstlisting}%
\caption{Type definitions of the trace structure.}
    \label{l:trace-syntax}
\end{figure}

We define a \emph{trace} data format that stores all necessary information about host function execution and their side effects.
By defining traces, we effectively decouple the record phase and replay phase and relay any required data by the fixed format of traces.

Listing \ref{l:trace-syntax} provides type definitions of the trace structure.
A \code{Trace} is a linear sequence of events.
\codens{Event}s correspond to units of behavior that happen during the 
Wasm app execution and that possibly involve interaction with host code.
There are six types of possible trace events: 
\code{FuncEntry}, \code{FuncReturn}, \code{Call}, \code{CallReturn}, \code{Load}, and \code{Store}.
The \code{funcidx} field in events corresponds to the function index of the Wasm function.
\code{ValType} corresponds to Wasm's four primitive types.
A \code{FuncEntry} event corresponds to the start of the function body.
This event represents the entrance to a Wasm-side function, 
with \code{params} storing the arguments given to the function call.
Conversely, a \code{FuncReturn} event corresponds to the end of the function body.
This event represents the exit from a Wasm-side function, 
with \code{values} storing the return values of the function.
\code{Call} and \code{CallReturn} events are produced by calling and returning from functions imported by a Wasm module.
Both events store the function index of the callee function, 
with the \code{CallReturn} event additionally storing the returned values.
A \code{Load} event corresponds to Wasm \wasms{load} instructions.
It stores the memory index in the \code{memidx} field, 
the address in the \code{address} field, and the loaded value in the \code{value} field.
Similarly, a \code{Store} event stores the memory index, address and stored value of Wasm \wasms{store} instructions.

In practice, we encode the trace in two formats: 
a textual format and a binary format.
The textual format is a JSON-like representation of the trace, 
where each trace event is encoded as an object, 
and the whole trace is encoded as a list of such objects.
For instance, a \code{Load} event may be represented as 
\code{Load \{ memidx: I32(0), address: I32(1000), value: I16(300) \}}.
We use the textual format for human-reading purposes, such as debugging.
In the binary format, each entry starts with a byte indicating the entry type, 
so the byte length of the entry is known.
We use the binary format in our implementation to reduce memory usage and for efficient parsing of traces.

\subsubsection{Instrumentation}
\label{s:instrumentation}

We use an instrumentation-based approach to record interactions between the Wasm module and the host environment and to store them into traces.
To this end, 
we add instructions to the Wasm binaries to capture runtime information, 
such as operand stack values.
Using instrumentation allows us to record traces from arbitrary 
Wasm-based web applications, without changing the web browser implementation 
or depending on features specific to certain platforms or libraries.

An important property of our instrumentation is that it should not change the original Wasm module's semantics.
Instead, the instrumented Wasm module serves as a drop-in replacement for the original Wasm module during recording, with the only behavioral change being the recording of a trace.
To preserve the Wasm module's semantics during instrumentation, 
our instrumentation strategy uses special recorder functions, which take runtime information as input parameters, 
records the event into the trace, and return.
We define recorder functions for each trace event.
For instance, the recorder function for a \code{load} instruction gets 
\code{memidx}, \code{address}, \code{value} as input arguments
and records the corresponding \code{Load} event.
Our instrumentation copies the runtime information on the stack, 
calls the imported recorder function, and then returns to the original execution flow.
By Wasm's function call semantics, 
calling the recorder function will consume only the copied values from the stack
and will not divert the original control flow.

\subsection{Reduce Phase}\label{s:trace_reduce}

The execution of a Wasm web application can produce millions of host interaction events.
A na\"ive approach would quickly run out of time and memory before a replay binary is generated. 
An essential component to make Wasm-R3 practical is to reduce the size of traces by filtering out events that contain redundant or  unnecessary information.
We call this process \emph{trace reduction}.
A key insight is that we only need to keep trace events related to non-determinism.
For instance, many of the \code{Store} events can be deterministically replayed by the original Wasm module itself.
Thus, we can conclude that \code{Store} events are not necessary to replay host-side non-determinism.

In this section, we describe two trace reduction techniques:
\emph{shadow memory optimization} (Section~\ref{s:shadow_opt}),
which filters out redundant \code{Store} and \code{Load} events, and 
\emph{call stack optimization} (Section~\ref{s:call-stack-opt}),
which filters out unnecessary \code{Call} and \code{CallReturn} events.

\subsubsection{Shadow Memory Optimization}
\label{s:shadow_opt}

Most of the \code{Store} and \code{Load} events in traces are not related to non-determinism and can thus be removed. 
It is not necessary to record all Wasm stores for accurate replay, as the original code performs the exact same sequence of stores as long as the current state of the program is the same.
Thus, it is only necessary to keep stores related to non-deterministic behavior, i.e., stores that come from the host.
Unfortunately, we cannot directly hook into host-side stores (e.g., when JavaScript writes into WebAssembly memory),
since we only instrument the application's Wasm code.
However, \code{Load} events on the Wasm side can observe when values in linear memory diverge from what was last recorded, which means they were modified by the host.

Inspired by memory optimization techniques in prior work~\cite{PinPlay,jalangi},
we apply the \emph{shadow optimization technique} to remove unnecessary \code{Store} and \code{Load} events and keep only \code{Load} events that observes the host-side side effect.
The technique maintains a data structure called \emph{shadow memory},
which keeps track of the written values to the original Wasm module's linear memory.
By comparing the loaded value of a \code{Load} event and the value stored in the shadow memory,
we can determine if the \code{Load} event observes the host-side side effect or not, and discard the unnecessary events.

\begin{figure}[t]
    \centering
    \includegraphics[width=\linewidth]{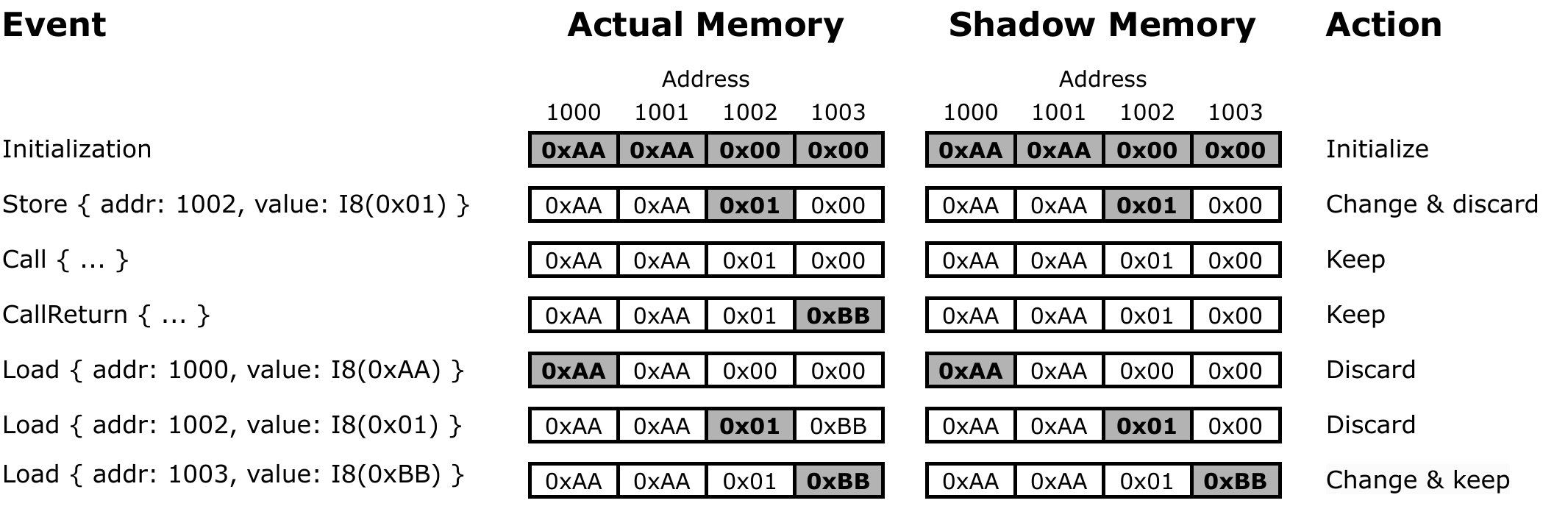}
    \caption{Example of the shadow memory optimization.}
    \label{f:shadow-mem}
\end{figure}

Figure~\ref{f:shadow-mem} illustrates how the shadow memory optimization works.
For each step of a Wasm module's execution,
we illustrate the corresponding trace event, the state of the module's
actual memory, and the state of the shadow memory.
For presentation brevity, we omit the irrelevant fields in the trace events.
In the actual memory states, we represent the parts that are read or written as gray cells.
In the shadow memory states, we represent the parts that are modified or compared with the loaded values as gray cells.

We first initialize the shadow memory to contain the same values as the original Wasm module,
by following the Wasm module's data section definition.
The first trace event is a \code{Store} of a single-byte value \code{0x01} to address \code{1002}.
The shadow memory optimization applies the same write to the shadow memory and discards the event.
Then, as a result of the call of an imported function in the second and third events,
the content at address \code{1003} is mutated.
The fourth event is a \code{Load} of value \code{0xAA} from address \code{1000}.
The optimization first compares the loaded value with the value at the same address in the shadow memory.
Because the values are equal, the optimization discards the \code{Load} event.
Similarly, the optimization discards the fifth event.
The sixth event is a \code{Load} of value \code{0xBB} from address \code{1003}.
The optimization compares the loaded value and the shadow memory value.
Here, the values are different, i.e., the value has been mutated as a side effect of interacting with the host.
Our optimization updates the shadow memory value as the loaded value and keeps the \code{Load} event.
As a result of shadow memory optimization, only the second, third, and sixth trace events are remaining, whereas the other, unnecessary \code{Store} and \code{Load} events are discarded.

\subsubsection{Call Stack Optimization}
\label{s:call-stack-opt}
While we record every trace event related to function execution, a significant portion of them are unrelated to non-determinism.
During the execution of a Wasm module, there are three possible kinds of function calls: \emph{export calls}, \emph{import calls}, and \emph{internal calls}.
Export calls are function calls from the host-side code to functions exported by the Wasm module.
Import calls are function calls from the Wasm module to functions imported from the host environment.
Internal calls are function calls from a function in the Wasm module to another internal function.
Among these three kinds of calls, we do not need to keep track of internal calls as those can be deterministically replayed by the original Wasm code.
Thus, we can safely remove \code{FuncEntry}, \code{FuncReturn}, \code{Call}, and \code{CallReturn} events produced by internal calls.  
In addition, we can remove \code{FuncReturn} events produced by returning from export calls.
This is because the same return values can be deterministically replayed by the functions defined in the original Wasm module. 

We apply \emph{call stack optimization} to remove trace events produced by internal calls.
To distinguish if an event was produced by an internal call or not, we track
function calls in our own \emph{call kind stack}.
The call kind stack stores \code{INT} and \code{EXT} objects, which correspond to Wasm-internal contexts and host-side code contexts, respectively.
Similar to conventional function call stacks, our call kind stack keeps track of calling contexts of trace events.
By observing the stack top element and the next trace event, we can determine if the event was produced by an internal call and discard it.

\begin{figure}[t]
  \centering
  \includegraphics[width=.85\linewidth]{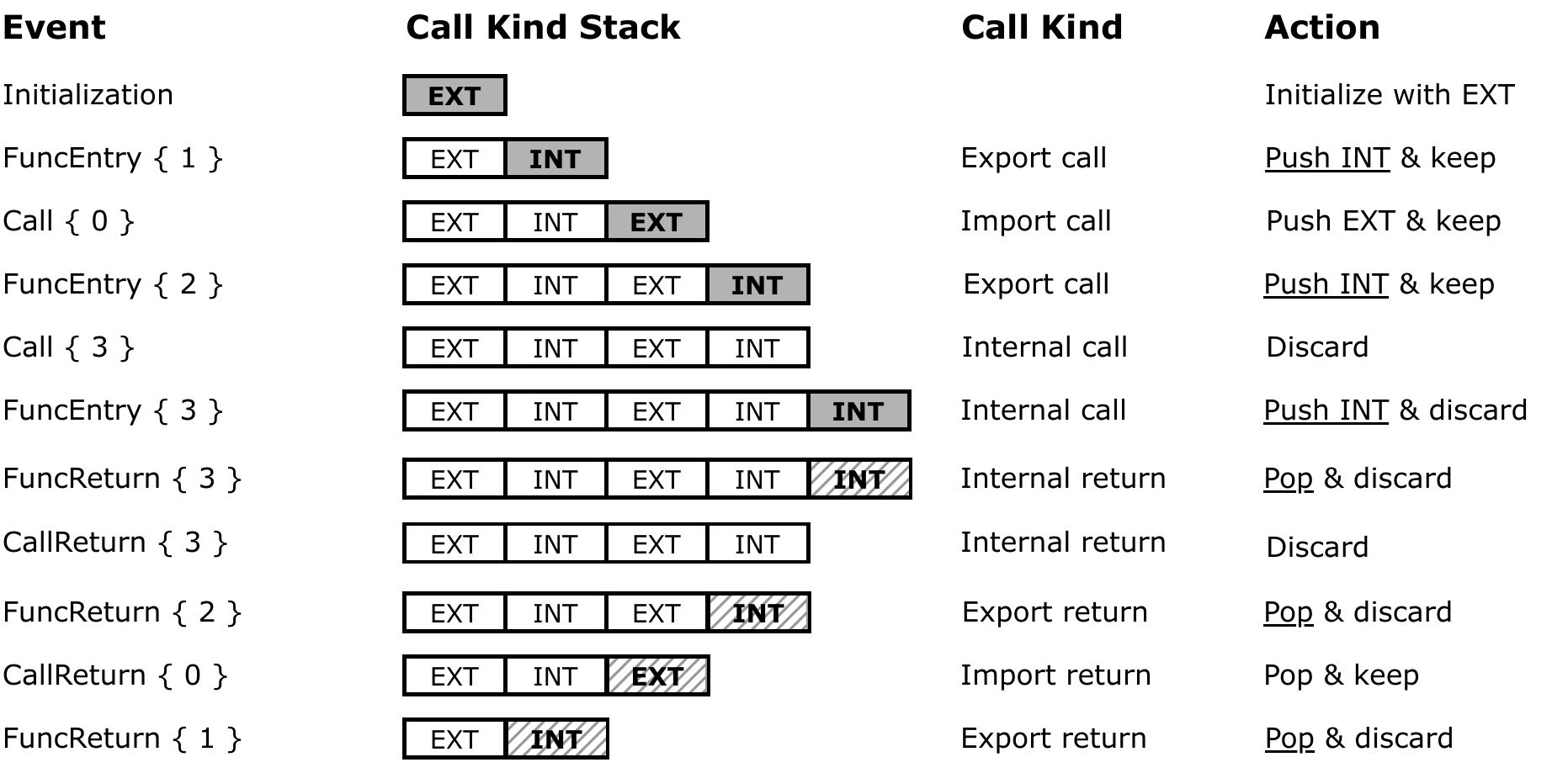}
  \caption{Example of the call stack optimization.}
  \label{f:call-stack}
\end{figure}

Figure \ref{f:call-stack} illustrates our call stack optimization process.
In this example, we assume that the function at index \code{0} is the only function imported from the host environment, and any other functions are internal functions of the original Wasm module.
We represent trace events by omitting unnecesary fields and only annotating the kind of an event and the \code{funcidx} field as a number.
The call kind stack column represents the state of the call kind stack at each step.
Gray-colored cells represent objects pushed in the current step, and gray-hatched cells represent objects popped in the current step.
The call kind column represents the kind of the function call produced by the corresponding trace event and whether the function is being called or returning.
The rightmost column shows which action the call stack optimization performs at each step. 
The underlined actions are always performed regardless of the event's \code{funcidx} or the call kind stack top content.
Note that we omitted \code{Store} or \code{Load} events from the trace for simplicity.

The call stack optimization rules are as follows.
We first initialize the call kind stack with a single \code{EXT} object.
This represents the outermost host execution context, e.g., JavaScript code which might call into exported Wasm functions. 
Then, we iterate over trace events.
On a \code{FuncEntry} event, we first observe the stack top.
If the top is \code{EXT}, then the event was produced by an export call; we keep the event.
If the top is \code{INT}, then the event was produced by an internal call; we discard the event.
Finally, we always push \code{INT} on the stack.
On a \code{Call} event, we first check the \code{funcidx} field.
If the index corresponds to an imported function, then the event was produced by an import call; we push \code{EXT} on the stack and keep the event.
If the index corresponds to an internal function, then the event was produced by an internal call; we discard the event. 
On \code{FuncReturn} event, we first pop an object from call kind stack and discard the event.
On a \code{CallReturn} event, we first check the \code{funcidx} field.
If the index corresponds to an imported function, then the event was produced by returning from an import call; we pop an object from the stack and keep the event.
If the index corresponds to an internal function, then the event was produced by returning from an internal call; we discard the event.

As illustrated in Figure~\ref{f:call-stack}, the call stack optimization discards all events produced by internal calls.
In addition, it discards all \code{FuncReturn} events produced by returning from export calls.
In the example, this filters out 6 out of 10 events.
The effect on real-world traces is evaluated in Section~\ref{s:evaluation}.

\subsection{Replay Phase}
\label{s:replay}
In the replay phase, Wasm-R3 gets an input trace and generates an excutable, standalone replay benchmark. 
In this process, Wasm-R3 uses a \emph{replay intermediate representation}, or \emph{replay IR}.
We describe the definition of the replay IR in Section~\ref{s:replay-ir}.
By first generating a replay IR (Section~\ref{s:trace2ir}) from a trace, we can apply the replay IR optimizations (Section~\ref{s:replay-opt}) to reduce the size of the replay IR.
Finally, Wasm-R3 translates this replay IR into one of three different output formats (Section~\ref{s:replay_outputs}) and generates a replay binary (Section~\ref{s:rir_to_benchmarks}).

\subsubsection{Replay IR}
\label{s:replay-ir}
Replay IR is a format designed to represent behaviors of \emph{replay functions}.
Replay functions are functions that implement the replay mechanism by replaying the return values and side-effects of host-side functions recorded during the record phase.
The goal of the replay phase is to generate a replay function for each function imported from the original Wasm module.
We utilize the replay IR as a general format that describes the behaviors of replay functions without depending on specific output formats.
Thus, introducing the replay IR effectively divides the problem of generating replay binaries for multiple output formats into three problems: 1) translating a trace into a replay IR, 2) optimizing the replay IR, and 3) translating the replay IR to each output format. 

\begin{figure}[t]
\begin{lstlisting}[language=Definition]
Action      := ExportCall | MutateMem
ExportCall  := { idx: I32, vals: Seq<ValType> }
MutateMem   := { idx: I32, addr: I32, val: I8 }
ValType     := I32 | I64 | f32 | f64
Context     := Seq<Action>
Function    := Seq<Context>
Replay      := Map<I32, Function>
\end{lstlisting}
\vspace*{-8pt}
\caption{Type definitions of the replay IR.}
\label{l:replay-ir}
\end{figure}

We present the definition of the replay IR in Figure~\ref{l:replay-ir}. 
An \code{Action} corresponds to a single instruction in the host: 
\code{ExportCall} or \code{MutateMem}.
An \code{ExportCall} represents a function call from the host to an exported Wasm function. 
A \code{MutateMem} represents a host-side effect of mutating the Wasm module's linear memory content.
A \code{Context} is a sequence of actions, which represents the actions executed during the context of a single function call.
Then, we define a \code{Function}, which is a sequence of contexts. 
Each \code{Context} corresponds to the instructions executed by $i$-th invocation of a host-side function, where $i$ is the index of the \code{Context} in the \code{Function} sequence.
Finally, we define \code{Replay} as a mapping from \code{I32} numbers to \codens{Function}s;
the \code{I32} numbers represent the function indices of the functions imported by the original Wasm module, and the mapped \codens{Function}s represent the corresponding replay functions.

\subsubsection{From Trace to Replay IR}
\label{s:trace2ir}
\begin{figure}[t]
\begin{lstlisting}
function translate(optimized_trace: Trace) -> (Replay, Function): 
  let GLOBAL_CONTEXT: Context = new Context()
  let ENTRY_FUNC: Function = new Seq(GLOBAL_CONTEXT) 
  let last_context: Ref<Context> = &GLOBAL_CONTEXT
  let context_stack: Stack<Ref<Context>> = Stack().push(&GLOBAL_CONTEXT)
  let replay: Replay = Map() 
  foreach event in optimized_trace:
    switch event:
      case FuncEntry: 
        context_stack.top().push(new ExportCall(event))
      case Call:
        let new_context = new Context()
        replay.get(event.funcidx).append(&new_context)
        context_stack.push(&new_context)
        last_context = &new_context
      case CallReturn:
        last_context = context_stack.pop()
      case Load: 
        let new_action = new MutateMem(event)
        if typeof last_context.last() == ExportCall:
          last_context.splice(last_context.length - 1, new_action)
        else:
          last_context.append(new_action)    

  return (replay, ENTRY_FUNC) 
\end{lstlisting}
\vspace*{-8pt}
\caption{Trace to replay IR translation algorithm.}
\label{f:translate-alg}
\end{figure}

We present the algorithm to translate a trace into a replay IR in Figure~\ref{f:translate-alg}.
The algorithm gets an input trace and returns a \code{Replay} and a \code{Function}.
The \code{Replay} value corresponds to the map from the imported function indices to the replay functions.
The \code{Function} value corresponds to the entry function to the Wasm module; as it does not correspond to a function index of the original Wasm module, we return it separately from the \code{Replay} value.

We use five variables in the algorithm: \code{GLOBAL\_CONTEXT}, \code{ENTRY\_FUNC}, \code{last\_context}, \code{context\_stack}, and \code{replay}. 
The variable \code{ENTRY\_FUNC} denotes the entry \code{Function}. 
Because the entry function is executed exactly once, it has only one \code{Context}, which is the variable \code{GLOBAL\_CONTEXT}. 
The variable \code{last\_context} is a reference to a context; it points to the latest context that new \code{MutateMem} actions should be appended to. 
The variable \code{context\_stack} is a stack of references to contexts.
We maintain \code{context\_stack} to keep track of the current host-side function call and push actions into correct contexts.  
We initialize \code{last\_context} and \code{context\_stack} with \code{GLOBAL\_CONTEXT} as the entry function is the first function executed in the whole Wasm web application execution.
Finally, the variable \code{replay} stores the \code{Replay} object which will be returned.
We assume that each \code{Function} object inside the \code{replay} is properly initialized with an empty sequence when we access it.

The goal of the algorithm is to fill the \codens{Function}s in the \code{replay} with \codens{Context}s and fill the \codens{Context}s, including \code{GLOBAL\_CONTEXT}, with \codens{Action}s, so that the return values correctly mirror the host-side behaviors recorded in the input trace.
To correctly mirror the input trace, 1) a correct \code{Action} must be created for each event and 2) the created \code{Action} must be inserted to a correct \code{Context}.
To do this, the algorithm uses case analysis on each event and properly updates \code{last\_context} and \code{context\_stack}. 
For a \code{FuncEntry} event, it creates an \code{ExportCall} action and pushes it to the \code{Context} of the current function.
We find the \code{Context} of the current function by referring to the top of \code{context\_stack} in this case.
For a \code{Call} event, it creates a new \code{Context} object, namely \code{new\_context}, for this import call.
This new \code{Context} object is appended to the \code{Function} at the \code{event.funcidx} index of \code{replay}.
Then, it pushes the reference to \code{new\_context} onto \code{context\_stack} and updates \code{last\_context}. 
For a \code{CallReturn} event, the current function call is returned, so it pops from \code{context\_stack}.
Then, by updating \code{last\_context} to the popped \code{Context} object, when we append \code{MutateMem} objects to \code{last\_context}, it correctly mirrors the fact that the side effect observed by the Wasm-side execution is caused by the most recently returned host-side execution.  
Lastly, for a \code{Load} event, it creates a new \code{MutateMem} action reflecting the side effect on the memory. 
Then, to insert this action into the correct position, it inspects the last element of \code{last\_context}. 
In most of the cases, it appends the new \code{MutateMem} at the last position of \code{last\_context}.
However, if the last action on the \code{last\_context} is an \code{ExportCall} action, this means that the mutation happened before the corresponding export call and the side effect was later observed.
Thus, it inserts the new \code{MutateMem} action right before the \code{ExportCall} action.

\subsubsection{Replay IR Optimizations}
\label{s:replay-opt}

We apply optimizations on replay IRs to reduce their size.
Before we generate an output replay benchmark, we reduce the replay IR size so that the size of the replay binary is also reduced.
By reducing the size of the replay binary, we address three practical issues.
\begin{itemize}
\item First, we produce valid replay binaries that are accepted by web browser engines.
Production web engines impose a common restriction on the size of functions inside the Wasm binary\footnote{See, e.g., \url{https://github.com/v8/v8/blob/master/src/wasm/wasm-limits.h}.}.
By na\"ively translating \code{Function} objects in a replay IR into Wasm functions, the size of replay functions may exceed the function size limit. 
To solve this issue, we employ replay IR optimizations to reduce the number of instructions of each replay function and produce smaller Wasm replay functions.
\item Second, we shorten the compilation time of the replay binaries, which correlates with the runtime performance of the replay benchmark.
A previous study \cite{wizard} has found out that the compilation time of Wasm modules significantly impacts the performance evaluation of a Wasm engine.
Since our replay IR optimization techniques reduce the overall size of the replay binary, we can reduce the impact of compilation time on the performance evaluation of Wasm engines.
\item Third, we reduce the execution time of the replay benchmarks by reducing the number of instructions in the final replay function bodies.
We achieve this effect by merging mulitple instructions into a single instruction while preserving the function behavior.
\end{itemize}

We describe our two replay IR optimization techniques: \emph{memory write merge} and \emph{function split}.

\paragraph*{Memory Write Merge Optimization}
By inspecting replay IRs, we found that Wasm often involves writing data to consecutive bytes in the Wasm linear memory.
Inspired by the \wasms{memory.init} instruction, which executes bulk writes on multiple bytes, we design the \emph{memory write merge} optimization that merges multiple \code{MutateMem} actions on consecutive addresses in a single action.
To represent the bulk write, we introduce a new action \code{BulkMutateMem}.
The \code{BulkMutateMem} action has the same syntax as \code{MutateMem}, except that the \code{val} field is a number type with no size limit. 
In theory, the memory write merge optimization can merge an unlimited number of \code{MutateMem} actions into a single \code{BulkMutateMem} action.
We translate the \code{BulkMutateMem} action into the \wasms{memory.init} Wasm instruction.

\begin{figure}[t]
\begin{subfigure}[t]{0.5\textwidth}
\begin{lstlisting}[language=Java,numbers=left,label=lst:null1,basicstyle=\scriptsize\ttfamily]
MutateMem {idx: 0, addr: 0, val: \08}
MutateMem {idx: 0, addr: 1, val: \07}
MutateMem {idx: 0, addr: 2, val: \06}
MutateMem {idx: 0, addr: 3, val: \05}
MutateMem {idx: 0, addr: 4, val: \04}
MutateMem {idx: 0, addr: 5, val: \03}
MutateMem {idx: 0, addr: 6, val: \02}
MutateMem {idx: 0, addr: 7, val: \01}
MutateMem {idx: 0, addr: 8, val: \00}
\end{lstlisting}
  \caption{Replay IR before the optimization.}
\end{subfigure}
~
\begin{subfigure}[t]{0.5\textwidth}
\begin{lstlisting}[language=Java,numbers=left,label=lst:null2,basicstyle=\scriptsize\ttfamily]
BulkMutateMem {
  idx: 0, 
  addr: 0, 
  val: "\08\07\06\05\04\03\02\01\00"
}
\end{lstlisting}
  \caption{Replay IR after the optimization.}
\end{subfigure}
  \caption{Memory write merge optimization example.}
  \label{f:mem-merge}
\end{figure}

Figure~\ref{f:mem-merge} illustrates an example of applying the memory write merge optimization.
The left-side figure is a part of the replay IR before applying the optimization.
It is a sequence of eight \code{MutateMem} actions on consecutive bytes.
The optimization merges the eight seperate writes in eight \code{MutateMem} actions into a single, 8-byte value in the \code{BulkMutateMem} action.

\paragraph*{Function Split Optimization}
As mentioned earlier, production web browser engines impose a common restriction on the size of a single Wasm function. 
We observed that some \code{Function} objects in replay IR exceed that maximum size limit.
These \code{Function} objects represent functions frequently called in the Wasm web application execution.
For instance, we observed a utility function that performs a conversion of \code{float64} values to integers corresponding to millions of actions.

To prevent our output replay binaries from being invalid, we employ \emph{function split optimizations} on such large \code{Function} objects.
In the function split optimization, we outline some parts of a \code{Function} object into other \code{Function} objects and replace them with calls to the new \code{Function} objects.
By setting an appropriate threshold of the \code{Function} object size, we maintain each \code{Function} object size below the maximum Wasm function size limit imposed by the web engines. 

\subsubsection{Output Formats}
\label{s:replay_outputs}
The replay benchmarks generated by Wasm-R3 in the replay phase consist of three parts.
First, the original Wasm module of the application.
Second, the \emph{replay code}, which replicates the behavior of the host environment during the record phase.
Third, the \emph{setup and instantiation code}, which links the first two parts together.
The setup and instantiation code fulfills the imports of the original Wasm module with functions from the replay code and starts the replay benchmark.
The last two parts are generated from the replay IR. 

While the original application code is always in Wasm, we can make different choices for the replay code and the setup and instantiation code.
If the replay code is in JavaScript, the setup and instantiation code should also be in JavaScript (option ``JS").
When the replay code is in Wasm, we can either statically link the replay code with the original Wasm module into a single, self-contained binary (option ``self-contained Wasm"), or generate JavaScript setup and instantiation code that loads the replay code and the original Wams module separately (option ``dynamic linking"). 
In the dynamic linking option, the actual linking between the original Wasm module and the replay code happens during the instantiation at runtime.

All three output formats are sensible choices. 
A self-contained Wasm replay is the easiest option for downstream consumers of the benchmark, as it can be executed in any Wasm engine, including Wasm standalone runtimes that do not have a JavaScript host environment.
We also expect a self-contained Wasm is the most performant way for replay because it does not involve function calls across the Wasm module and JavaScript host environments. 
Hence, we choose the self-contained Wasm format by default and use it for evaluation in Section \ref{s:evaluation}.
However, when the replay code is kept in a separate Wasm module or in JavaScript, this can be useful to benchmark cross-language or multi-module interactions.
In Wasm-R3, we provide options to select from three output formats, so users can generate replay benchmarks according to their use cases.

\subsubsection{From Replay IR to Output Formats}
\label{s:rir_to_benchmarks}

Once we have a replay IR, the generation of an executable, standalone replay benchmark is carried out in a straightforward, single-pass manner.
For each \code{Function} in the replay IR, we generate a replay function according to the output mode.
For JavaScript, this would be a JavaScript function, and for Wasm, this would be a Wasm function.
We then define a global counter variable for each function to keep track of the current \code{Context}.
The body of the generated function consists of a switch statement in the respective language that maps different counter values to different sequences of instructions, followed by a part that increments the counter for each invocation.
Each sequence of instructions that are translated from a \code{Context} is
a series of simple line-by-line translations of \codens{Action}s in the replay IR to their
corresponding instructions in the langauge.
For example, \codens{MutateMem}s are translated to their corresponding \wasms{store} instructions.

\section{Implementation}
\label{sec:implementation}
In this section, we describe notable implementation details of Wasm-R3.

\paragraph*{Implementation Summary.}
The implementation of Wasm-R3 amounts to roughly 2,200 lines of TypeScript and 2,200 lines of Rust, divided into the frontend (the record and reduce phases) and the backend (the replay phase).
The code is released under the MIT License and is publicly available at \url{https://github.com/sola-st/wasm-r3}.
Wasm-R3 utilizes two third-party libraries: Wasabi~\cite{Wasabi}, an instrumentation framework for Wasm, which we use to inject calls to recorder functions and to store corresponding trace events, and Binaryen~\cite{Binaryen}, a compiler toolchain and infrastructure, which we use to optimize replay binaries.

\paragraph*{Proxy.}
Rather than intrusively modify web browsers (or Wasm engines in the web browsers), we employ a \emph{proxy} to intercept Wasm and JavaScript code.
Modern web browsers expose the capability to intercept and modify network requests and responses.
For instance, Chromium provides the DevTools Protocol~\cite{devtools} for this purpose.
The proxy component leverages this capability to intercept JavaScript files and patch\footnote{Of course, monkey-patching JavaScript has robustness issues, and some modules can be missed.} the function definitions of Wasm module instantiation APIs: \code{WebAssembly.Instance}, \code{WebAssembly.instantiate}, and \code{WebAssembly.instantiateStreaming}.
The new instantiation functions intercept the Wasm binary before instantiation and inject instrumentation before forwarding them to the underlying Wasm engine.
We utilize the Playwright library \cite{playwright} to implement the proxy across all major web browsers.
This proxy approach allows Wasm-R3 to instrument every Wasm module on-the-fly without modifying the browser implementation.

\paragraph*{Mutable State in Wasm.}
In Wasm, there are three kinds of mutable states in instances: globals, tables, and memories.
Globals are mutable or immutable, single-value storage that can be imported and exported from Wasm modules.
Tables store opaque references to host objects and Wasm functions.
Since the \wasms{call\_indirect} instruction indirects through a table, mutable tables can be used to implement dynamic linking via indirect calls.
Like Wasm linear memory, globals and tables can be modified by the host environment if exported from an instance.
Thus, in order to replay all nondeterminism from the host, we also need to record mutation of globals and tables. 
Although not described in detail here, Wasm-R3 also records and replays global and table mutations by instrumenting \wasms{global.get} and \wasms{table.get} instructions, respectively.
Similarly, it also employs shadow memory optimizations for globals and tables to distinguish mutations from the host environment and the module itself.

\paragraph*{Simultaneous Record and Reduce.}
In our implementation of Wasm-R3, the reduce phase is partially overlapped with the recording phase; the reduce phase filters out most redundant trace events even before they are stored in a trace.
This is done primarily by the shadow memory optimization and the call stack optimization algorithms, which are applied online in the recorder functions.
For example, a \wasms{load} instruction simply reads both the shadow and real memories and suppresses generating a load event if the two values are the same, which implies that the program either writes the value or has already observed a host-written value.
Similarly, the call stack optimization algorithm is included directly in the recorder function.
Filtering events requires more checks, but is less expensive than generating events and then later filtering them out, which naturally saves space but also reduces the overall recording overhead.

\section{Evaluation}
\label{s:evaluation}
We evaluate Wasm-R3 by addressing the following four research questions.

\begin{itemize}[labelindent=\parindent,leftmargin=*]
  \item \textbf{RQ1. Applicability}:
  To what extent does Wasm-R3 apply to real-world web applications and different Wasm engines?
  \item \textbf{RQ2. Performance}:
  How much overhead does Wasm-R3's record phase introduce?
  What are the performance characteristic of the replay benchmarks?
  \item \textbf{RQ3. Effectiveness of trace reduction}:
  To what extent do our trace reduction techniques reduce the size of the recorded traces?
  \item \textbf{RQ4. Effectiveness of replay optimization}:
  By how much do our replay optimization techniques reduce the size of replay binaries? How do the optimizations impact the performance characteristics of the replay benchmarks?
\end{itemize}

\subsection{Experimental Setup}
\label{s:setup}


We collect URLs of real-world \emph{Wasm web applications}, which we define to be interactive webpages that load at least one Wasm module, to serve as our evaluation targets.
To find such applications, we use two websites as starting points:
\textit{Made with WebAssembly} \cite{madewithwebassembly}, which is an open-source website that showcases projects created with Wasm, and
\textit{Awesome-Wasm} \cite{awesome-wasm}, an open-source repository that lists Wasm-related webpages.
We gathered URLs for Wasm web applications from these webpages by first manually crawling them and their subpages and filter out inaccessible websites, e.g. 404 or that require authentication, non-interactive webpages, and pages where the relevant Wasm APIs, e.g. \code{WebAssembly.Instance}, \code{WebAssembly.instantiate} or \code{WebAssembly.instantiateStreaming}, are not used in any script.

For repeatability, we use test scripts that automatically interact with each website using the Playwright library \cite{playwright} by mimicking common use cases.
Each script executes multiple user actions, such as clicking a button or typing text into an input form.
For websites that use multiple Wasm modules, we write multiple test scripts, each focusing on a different module.
These scripts run against live sites, which of course evolve over time and automation breaks.
In fact, the difficulties of reliably and reproducibly automating the execution of Wasm web applications is a key motivation of Wasm-R3, which ultimately produces completely self-contained Wasm benchmarks.

While writing test automation scripts, we exclude some applications from our evaluation targets that have one or more of the following problems:

\begin{enumerate}
  \item Throw errors even when used without Wasm-R3.
  \item Require external files or privileged information. 
    For example, we remove a GameBoy emulator \cite{binjgb} because it requires a GameBoy ROM file that we could not supply.
    \item Take unreasonably long time to download required data from the network. 
      This affects several video game ports, such as Arxwasm \cite{arxwasm} and D3wasm \cite{d3wasm}.
    \item Though statically appear to use Wasm, don't dynamically load any Wasm modules. 
      For example, we exclude Wasmboy \cite{wasmboy} because we could not automate it to load Wasm.
    \item Require automation scripts that would interact with the HTML canvas element. 
As the Playwright library does not provide APIs to recognize images inside a canvas element, we cannot perform any meaningful interactions with such applications beyond randomly clicking inside the canvas.
    \item Exhibit flakiness in the automation scripts without any meaningful errors. 
      For example, we exclude the application a tic-tac-toe game \cite{tic-tac-toe}, because it sometimes fails to load the game.
\end{enumerate}

\begin{table}[t]
\renewcommand{\baselinestretch}{1.08}
\centering
\caption{List of evaluation target Wasm web applications.}
\scriptsize 
\label{t:eval-target}
\vspace*{-1em}
\begin{tabular}{@{}l|l|l|c@{}}
\hline
\multicolumn{1}{c|}{\bf Name} &
\multicolumn{1}{c|}{\bf URL} &
\multicolumn{1}{c|}{\bf Domain} &
\multicolumn{1}{c}{\bf Success} \\
   \hline
\name{boa}             & \url{https://boajs.dev/boa/playground}                                               & Progr. lang.     & \cmark   \\
\name{bullet}          & \url{https://magnum.graphics/showcase/bullet}                                        & Simulator        & \cmark   \\
\name{commanderkeen}   & \url{https://www.jamesfmackenzie.com/chocolatekeen}                                   & Video game       & \cmark   \\
\name{factorial}       & \url{https://www.hellorust.com/demos/factorial/index.html}                            & Mathematics      & \cmark   \\
\name{ffmpeg}          & \url{https://w3reality.github.io/async-thread-worker/examples/wasm-ffmpeg/index.html} & Media            & \cmark   \\
\name{fib}             & \url{https://takahirox.github.io/WebAssembly-benchmark/tests/fib.html}                & Benchmark        & \cmark   \\
\name{figma-startpage} & \url{https://www.figma.com}                                                           & Graphics         & \cmark   \\
\name{fractals}        & \url{https://raw-wasm.pages.dev}                                                     & Graphics         &         \\
\name{funky-kart}      & \url{https://www.funkykarts.rocks/demo.html}                                          & Video game       & \cmark   \\
\name{game-of-life}    & \url{https://playgameoflife.com}                                                     & Video game       & \cmark   \\
\name{gotemplate}      & \url{https://gotemplate.io}                                                          & Progr. lang.     &         \\
\name{guiicons}        & \url{https://raylibtech.itch.io/rguiicons}                                            & Utility          & \cmark   \\
\name{hnset-bench}     & \url{https://raw.githack.com/gorhill/uBlock/master/docs/tests/hnset-benchmark.html}   & Benchmark        &         \\
\name{hydro}           & \url{https://cselab.github.io/aphros/wasm/hydro.html}                                 & Simulator        & \cmark   \\
\name{image-convolute} & \url{https://takahirox.github.io/WebAssembly-benchmark/tests/imageConvolute.html}     & Benchmark        &         \\
\name{jqkungfu}        & \url{http://jqkungfu.com}                                                            & Progr. lang.     & \cmark   \\
\name{jsc}             & \url{https://mbbill.github.io/JSC.js/demo/index.html}                                 & Progr. lang.     & \cmark   \\
\name{lichess}         & \url{https://lichess.org/analysis}                                                    & Video game       &         \\
\name{livesplit}       & \url{https://one.livesplit.org}                                                      & Utility          &         \\
\name{mandelbrot}      & \url{http://whealy.com/Rust/mandelbrot.html}                                          & Graphics         & \cmark   \\
\name{multiplyDouble}  & \url{https://takahirox.github.io/WebAssembly-benchmark/tests/multiplyDouble.html}     & Benchmark        & \cmark   \\
\name{multiplyInt}     & \url{https://takahirox.github.io/WebAssembly-benchmark/tests/multiplyInt.html}        & Benchmark        & \cmark   \\
\name{ogv}             & \url{https://brionv.com/misc/ogv.js/demo}                                            & Media            &         \\
\name{onnxjs}          & \url{https://microsoft.github.io/onnxjs-demo/\#}                                      & ML &         \\
\name{pacalc}          & \url{http://whealy.com/acoustics/PA\_Calculator/index.html}                            & Mathematics      & \cmark   \\
\name{parquet}         & \url{https://google.github.io/filament/webgl/parquet.html}                            & Graphics         & \cmark   \\
\name{pathfinding}     & \url{https://jacobdeichert.github.io/wasm-astar}                                     & Benchmark        & \cmark   \\
\name{playnox}         & \url{https://playnox.xyz}                                                            & Video game       &         \\
\name{rfxgen}          & \url{https://raylibtech.itch.io/rfxgen}                                               & Utility          & \cmark   \\
\name{rguilayout}      & \url{https://raylibtech.itch.io/rguilayout}                                           & Utility          & \cmark   \\
\name{rguistyler}      & \url{https://raylibtech.itch.io/rguistyler}                                           & Utility          & \cmark   \\
\name{riconpacker}     & \url{https://raylibtech.itch.io/riconpacker}                                          & Utility          & \cmark   \\
\name{roslyn}          & \url{http://roslynquoter-wasm.platform.uno}                                          & Progr. lang.     &         \\
\name{rtexpacker}      & \url{https://raylibtech.itch.io/rtexpacker}                                           & Utility          & \cmark   \\
\name{rtexviewer}      & \url{https://raylibtech.itch.io/rtexviewer}                                           & Utility          & \cmark   \\
\name{rustpython}      & \url{https://rustpython.github.io/demo}                                              & Progr. lang.     &         \\
\name{sandspiel}       & \url{https://sandspiel.club}                                                         & Video game       & \cmark   \\
\name{sqlgui}          & \url{http://kripken.github.io/sql.js/examples/GUI}                                   & Progr. lang.     & \cmark   \\
\name{sqlpractice}     & \url{https://www.sql-practice.com}                                                    & Progr. lang.     &         \\
\name{takahirox}       & \url{https://takahirox.github.io/WebAssembly-benchmark}                              & Benchmark        &         \\
\name{timestretch}     & \url{https://superpowered.com/js-wasm-sdk/example\_timestretching}                    & Media            &         \\
\name{waforth}         & \url{https://el-tramo.be/waforth}                                                     & Progr. lang.     &         \\
\name{wheel}           & \url{https://boyan.io/wasm-wheel}                                                    & Benchmark        &         \\
\hline
  \end{tabular}
\vspace*{-1em}
\end{table}

As a result, we compiled 43 URLs to Wasm web applications.
Table \ref{t:eval-target} summarizes our evaluation targets.
Our evaluation targets are composed of
9 programming language applications,
8 Wasm benchmarks,
6 video games,
4 graphics applications,
3 media applications,
2 mathematical computation applications,
2 simulator applications, and
1 ML(machine learning) application.
We claim these represent real-world Wasm web applications as we gathered them from well-known, open-source compilations of Wasm web applications and include applications from various domains.


We evaluate the benchmarks created with Wasm-R3 on three web browser engines (SpiderMonkey~\cite{SpiderMonkey} version 125.0b7, V8~\cite{V8} version 12.5.149, and JavaScriptCore~\cite{JavaScriptCore} version 277039) and three standalone Wasm engines (Wizard~\cite{WizardEngine} version 24$\alpha$.1998, Wasmtime~\cite{Wasmtime} version 19.0.1, and Wasmer~\cite{Wasmer} version 4.2).
We use the standalone-Wasm output format of the benchmarks for the entire evaluation (Section~\ref{s:replay_outputs}), as they can be executed across both web browser engines and standalone Wasm engines.
To run a standalone-Wasm benchmark in a browser, we use a simple JavaScript wrapper that loads the replay benchmark and calls its entry function. 


Our experiments are conducted on a machine running Ubuntu 22.04.1, equipped with an Intel Core i9-13900k CPU and 192GB of DRAM. 
With the Intel Core i9-13900k, we disable E-cores and use only P-cores, and set the Linux CPU frequency governor to performance mode for consistent results. 
We use Chromium 123.0.6312.4 as the browser for the proxy component in the record phase.
For the experiments in RQ2 and RQ4, we repeat each measurement ten times for each target.

\subsection{RQ1. Applicability}

We evaluate the applicability of Wasm-R3 in two ways.
First, we evaluate its ability to produce accurate benchmarks from various real-world Wasm web applications, which we call the \textit{accuracy experiment} (Section~\ref{s:accuracy experiment}).
Second, we evaluate to what extent the produced benchmarks execute successfully across different Wasm engines, which we call the \textit{portability experiment} (Section~\ref{s:portability experiment}).

\subsubsection{Accuracy Experiment}
\label{s:accuracy experiment}

We evaluate how accurately Wasm-R3's replay benchmarks match their execution in Wasm web applications.
The term ``accurate'' here means that the original web application and the corresponding replay benchmarks show the same behavior.
We assess accuracy by recording traces of both executions and test if both traces are exactly the same. 

Table~\ref{t:eval-target} shows for each application whether we could successfully produce accurate replay benchmarks.
In total, Wasm-R3 produces accurate replay benchmarks for 27 out of 43 applications.
These applications cover a wide range of domains, including programming language applications, graphics applications, and video games.
To the best of our knowledge, the resulting set of benchmarks is the first executable benchmark suite of real-world Wasm web applications.
We refer to these benchmarks as \emph{Wasm-R3-Bench} and, unless mentioned otherwise, use them throughout the rest of the evaluation.

Due to the complexity of the \web{} and limitations of the libraries that Wasm-R3 uses, the approach fails to produce accurate benchmarks for the remaining 16 applications.
We categorize the failures into three groups:
\begin{itemize}[leftmargin=1.5em]
  \item \emph{Implementation limitations (5 cases)}.
  Some failures are due to the known limitations of our implementation.
  For \name{image-convolute}, the trace contains over a million calls to a single host function.
  The current function split optimization works only inside a single context in the replay IR, which prevents the application of the optimization in this case; \name{playnox} fails for the same reason.
  We believe this would be resolved with a more advanced function split optimization that works across multiple contexts.
  For \name{timestretch}, \name{hnset-bench}, and \name{wheel}, our proxy logic does not seem to work the use cases of the applications.
  \item \emph{Dependency limitations (4 cases)}.
  Some failures are caused by the limitations of the libraries that Wasm-R3 uses.
  Wasabi fails to instrument \name{fractals} and \name{lichess}, because they use the SIMD proposal, and \name{livesplit}, because it uses the threads proposal, which are not yet supported by Wasabi.
  Binaryen fails on \name{waforth} because the library does not support block-type parameters.\footnote{\url{https://github.com/WebAssembly/binaryen/issues/6407}}
  \item \emph{Unknown problems (7 cases)}.
  For the remaining 7 applications, we could not determine the cause of the failure. We are currently investigating their cause.
  
\end{itemize}

\subsubsection{Portability Experiment}
\label{s:portability experiment}

The following experiment evaluates to what extent the replay benchmarks generated by Wasm-R3 execute successfully across different Wasm engines.
We run the portability experiment with all 27 accurate replay benchmarks, trying to run them on three web browser engines and three Wasm standalone engines (Section~\ref{s:setup}).
When running the portability experiment with web browser engines, we experiment with different optimization tiers of each engine. 
We count the experiment as successful if the replay benchmark runs successfully on all optimization tiers of the engine.
Likewise, as the Wizard and Wasmer engines also provide different optimization tiers, we also experiment with them.
All replay benchmarks successfully run across all execution tiers of the three web browser engines and three Wasm standalone engines.
In summary, our experiments show that Wasm-R3 is applicable in various usage scenarios.
In particular, we produce a suite of 27 replay benchmarks from real-world Wasm web applications, and these benchmarks run successfully on various Wasm engines.

\subsection{RQ2. Performance}
We evaluate the performance of Wasm-R3 from two perspectives. 
First, we assess the amount of overhead introduced during the recording phase (Section~\ref{s:record overhead experiment}).
Keeping the overhead low is crucial to minimize disruption to user interactions during the recording phase.
Second, we examine the performance characteristics of the replay benchmarks by measuring the time spent in code of the original Wasm module and in code added by Wasm-R3 to enable replay (Section~\ref{s:replay characteristic experiment}).
For a replay binary to be useful for evaluating the performance of Wasm engines, the majority of time should be spent in the original Wasm code.

\subsubsection{Record Overhead Experiment}
\label{s:record overhead experiment}

\begin{figure}[t]
  \includegraphics[width=.75\linewidth]{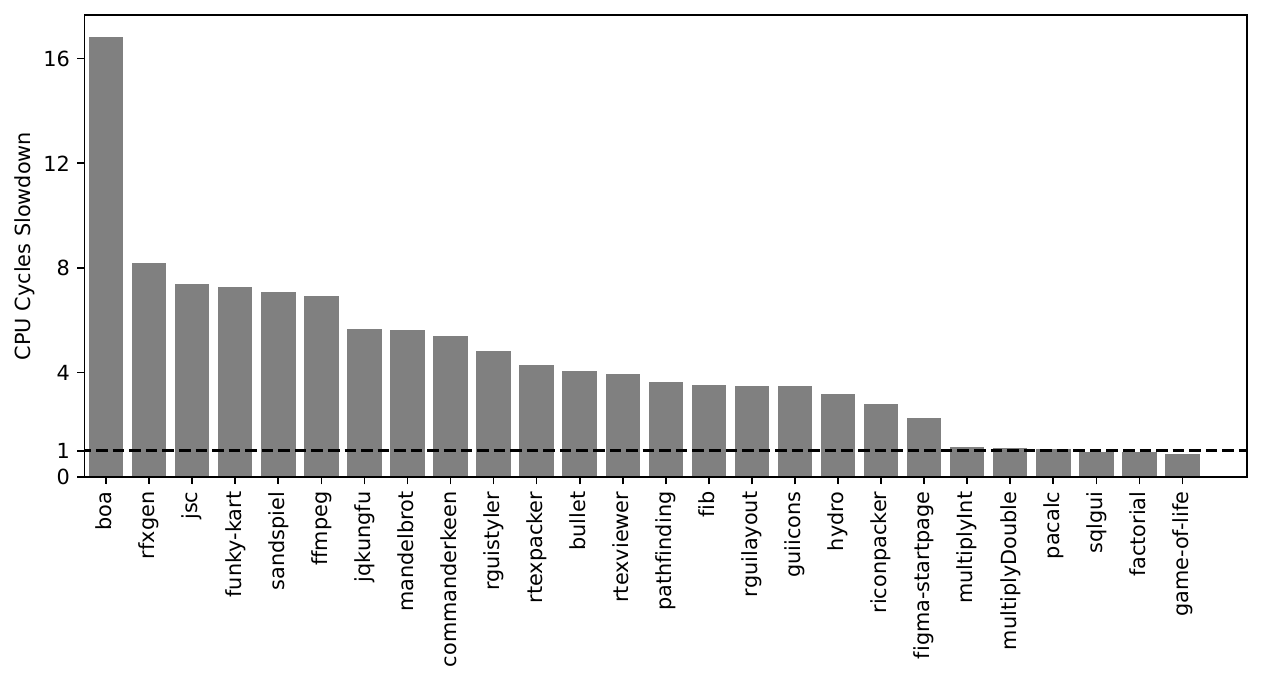}
   \caption{Relative increase in CPU cycles spent by the Chromium renderer process over uninstrumented application (2.0 = twice as many cycles, 1.0 = same).}
  \label{f:perf-record}
\end{figure}

The following experiment evaluates the overhead of Wasm-R3's record phase.
We measure the total CPU cycles spent by the Chromium renderer process of the target application using the Linux perf tool.
Among the 27 replay binaries in Wasm-R3-Bench, we exclude \name{parquet} in this experiment as our record overhead measurement infrastructure does not support its use of the WebGL library.
To compute the recording overhead, we compare the performance of the application when running with an uninstrumented and an instrumented Wasm module.
As in RQ1, we use our UI-level test scripts to simulate user interactions.
Because the test scripts are at the UI level, there is a risk of having slightly different workloads in different runs.
To mitigate this risk, and also to account for the inherent noise of performance measurements, we repeat each measurement across ten runs and compute the arithmetic mean of the spent CPU cycles.\footnote{For \name{fib}, we compute the arithmetic mean of nine runs only, because one recording run failed due to flakiness.} Then, we compute the ratio of the arithmetic mean of the CPU cycles spent by the instrumented application to the arithmetic mean of the CPU cycles spent by the uninstrumented application.

Figure~\ref{f:perf-record} shows the overhead introduced by the record phase of Wasm-R3.
We measured the CPU cycles spent executing JavaScript and Wasm code with and without instrumentation.
While the overhead varies across applications, the slowdown is generally modest, with a median of approximately 3.79\XX, a geometric mean of 3.40\XX, and all but one application exhibiting less than 8.18\XX overhead.
In practice, this overhead is acceptable; applications are still interactive and users can record realistic usage scenarios with Wasm-R3.

\subsubsection{Replay Characteristics Experiment}
\label{s:replay characteristic experiment}

We now evaluate the performance characteristics of the replay benchmarks created by Wasm-R3.
Wasm-R3 repackages the functions of the original Wasm module with replay functions, creating a standalone executable.
While any Wasm workload can serve as a benchmark, the overall goal is to capture the performance characteristics of the \emph{original} Wasm code.
A key metric is then the proportion of the work spent in the original functions versus the replay functions.
To answer this question, we first measure the CPU cycles per function using the \C{fprofile} monitor of Wizard.
Then, we distinguish whether the function belongs to the original Wasm module or the replay functions.
Summing the CPU cycles spent in each group, we can calculate the total cycles spent in the original Wasm module and the replay functions.
We repeat the experiment ten times to get the arithmetic mean value of the CPU cycles to reduce variance in the measurements.

\begin{figure}[t]
  \includegraphics[width=.8\linewidth]{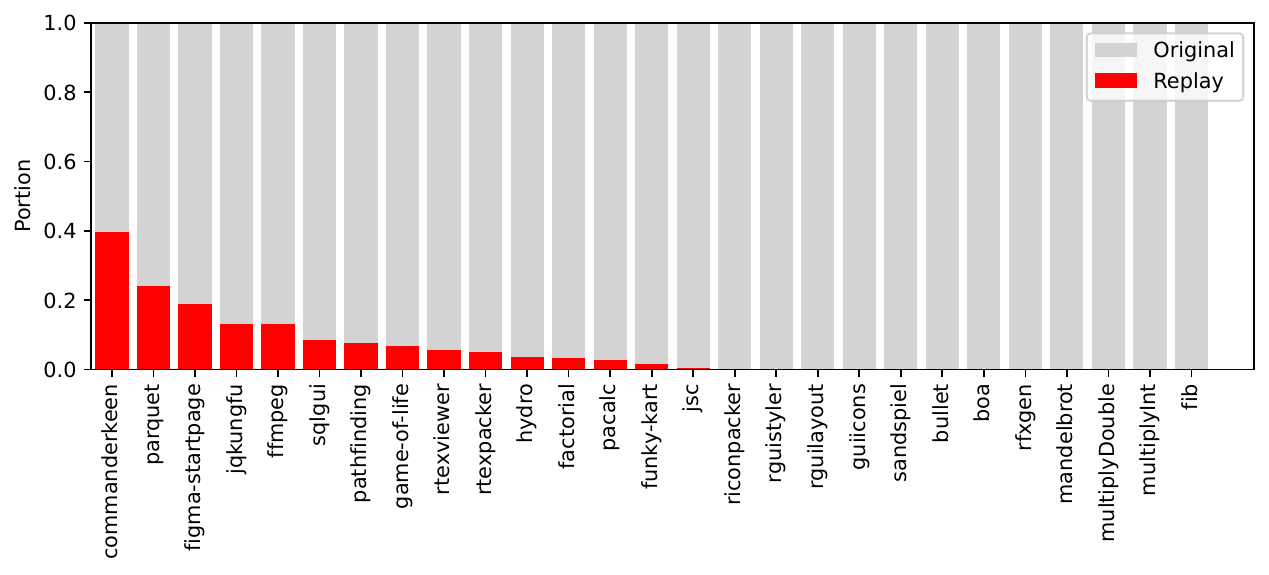}
  \caption{Proportion of the execution spent in the original Wasm modules (light
  gray) and replay functions (red), in CPU cycles.}
  \label{f:perf-replay}
\end{figure}

Figure \ref{f:perf-replay} displays the results of this experiment. 
The upper portions of the bars in light gray color represent the percentage of the cycles spent in the functions from the original Wasm module, while the lower portions in red color represent the percentage of the cycles spent in replay functions.
Ideally, a benchmark would spend 100\% of the cycles in the original Wasm code and 0\% in replay.
For the benchmarks we gathered, we find the geometric mean of the cycles spent in the replay binary to be 0.20\%.
Half of the benchmarks spend less than 1.53\%, all but one less than 24\%, and the maximum cycles spent in the replay part is 39\%.
The proportion varies according to the frequency and nature of the application's interactions with the host.
Yet, recall that our technique only requires the replay return values and observed side-effects (e.g. memory modifications) from host calls.
Side-effects from the host that are never observed by the application, as well as host functions that only read application memory, do not generate trace events.
This leads to the somewhat counter-intuitive result that applications that \emph{produce} a lot of data for the host (e.g. rendering frames for a game) actually do not generate many trace events; they generate (relatively small) trace events when the user interacts with the game.

Thus, in summary, our results show that the modules in Wasm-R3-Bench mostly exercise the behavior of the original, real-world applications, making these executions suitable for evaluating the performance of Wasm engines.

\subsection{RQ3. Effectiveness of Trace Reduction}

We evaluate how effective Wasm-R3's trace reduction techniques (Section~\ref{s:trace_reduce}) are in reducing the size of the recorded traces.
We record traces with four variants of the approach: 1) without any trace reductions, 2) only with the shadow memory optimization, 3) only with the call stack optimization, and 4) with both optimization techniques.
The fourth variant corresponds to the full Wasm-R3 approach.
We target all 27 modules in Wasm-R3-Bench, measure the size of the traces produced by the four variants, and report the differences in them.
By ``the size of a trace,'' we here mean the number of trace events recorded by Wasm-R3.

\begin{table}[t]
\footnotesize
  \caption{Trace reduction experiment results.}
  \label{t:table-reduce}
\begin{tabular}{l|r|r|r|r}
\hline
\multirow{2}{*}{\bf Name} &
  \multicolumn{4}{c}{\bf Trace (\# Events)} \\
\cline{2-5}
&
\multicolumn{1}{c|}{\bf No-opt} &
\multicolumn{1}{c|}{\bf Shadow-opt} &
\multicolumn{1}{c|}{\bf Call-stack-opt} &
\multicolumn{1}{c}{\bf All-opt} \\
\hline \hline
\name{boa}             & 1422316    & 41.36\%  & 58.65\%  & 0.01\%   \\ \hline
\name{bullet}          & 60731762   & 14.27\%  & 85.75\%  & 0.01\%   \\ \hline
\name{commanderkeen} & 190191109    & 21.12\%  & 79.12\%  & 0.23\%   \\ \hline
\name{factorial}       & 1483       & 35.87\%  & 66.01\%  & 1.89\%   \\ \hline
\name{ffmpeg}          & 83         & 36.14\%  & 87.95\%  & 24.10\%  \\ \hline
\name{fib}             & 4912039416 & 100.00\% & 0.00\%   & 0.00\%   \\ \hline
\name{figma-startpage} & 3065       & 61.01\%  & 54.71\%  & 15.73\%  \\ \hline
\name{funky-kart}      & 170064681  & 29.63\%  & 71.86\%  & 1.48\%   \\ \hline
\name{game-of-life}    & 2178       & 16.71\%  & 85.31\%  & 2.02\%   \\ \hline
\name{guiicons}        & 87593617   & 8.48\%   & 91.53\%  & 0.01\%   \\ \hline
\name{hydro}           & 4494       & 37.96\%  & 62.35\%  & 0.31\%   \\ \hline
\name{jqkungfu}        & 83         & 48.19\%  & 80.72\%  & 28.92\%  \\ \hline
\name{jsc}             & 1829171    & 69.12\%  & 35.63\%  & 4.75\%   \\ \hline
\name{mandelbrot}      & 151019238  & 0.90\%   & 99.21\%  & 0.11\%   \\ \hline
\name{multiplyDouble}  & 24         & 100.00\% & 100.00\% & 100.00\% \\ \hline
\name{multiplyInt}     & 24         & 100.00\% & 100.00\% & 100.00\% \\ \hline
\name{pacalc}          & 194039     & 43.05\%  & 58.95\%  & 2.00\%   \\ \hline
\name{parquet}         & 16736      & 54.92\%  & 60.40\%  & 15.32\%  \\ \hline
\name{pathfinding}     & 30798960   & 35.64\%  & 64.44\%  & 0.08\%   \\ \hline
\name{rfxgen}          & 143104877  & 6.79\%   & 93.21\%  & 0.00\%   \\ \hline
\name{rguilayout}      & 50654126   & 8.77\%   & 91.25\%  & 0.02\%   \\ \hline
\name{rguistyler}      & 63933807   & 8.62\%   & 91.40\%  & 0.02\%   \\ \hline
\name{riconpacker}     & 1078843    & 10.12\%  & 90.10\%  & 0.22\%   \\ \hline
\name{rtexpacker}      & 10         & 60.00\%  & 100.00\% & 60.00\%  \\ \hline
\name{rtexviewer}      & 10         & 60.00\%  & 100.00\% & 60.00\%  \\ \hline
\name{sandspiel}       & 300347531  & 18.65\%  & 81.38\%  & 0.03\%   \\ \hline
\name{sqlgui}          & 250114     & 39.27\%  & 63.05\%  & 2.32\%   \\ \hline \hline
\textbf{Geomean}  &           & \textbf{27.20\%}  & \textbf{38.17\%}  & \textbf{0.47\%} \\ \hline
\end{tabular}

\end{table}

Table~\ref{t:table-reduce} shows the results.
Both trace reduction techniques are effective in reducing the trace size.
On average, the shadow memory optimization and the call stack optimization reduce traces to 27.20\% and 38.17\% of the original trace size, respectively.
Together, the two optimizations reduce the traces to only 0.47\% of the original size, i.e. a more than 200\XX reduction.
In the development process, before optimization, Wasm-R3 initially failed to produce traces for most applications due to the lack of memory or timeouts.
However, after applying our trace reduction techniques, Wasm-R3 filters out a large portion of trace events and succeeds to produce traces for real-world applications.
Hence, we believe that our trace reduction techniques are essential to produce replay benchmarks from real-world Wasm web applications.

From the table, three interesting cases emerge: \name{multiplyDouble}, \name{multiplyInt}, and \name{fib}. 
For \name{multiplyDouble} and \name{multiplyInt}, the optimizations do not remove any trace events.
This is because \name{multiplyDouble} and \name{multiplyInt} do not perform any loads or calls during their execution, i.e., there is nothing for our techniques to optimize.
Instead, all the events in the traces are \code{FuncEntry} events directly followed by \code{FuncExit} events. 
These functions use a loop internally within an exported function to repeat the multiplications. 
In contrast, the results for \name{fib} show that almost all trace events are filtered out, with only 24 out of 4.9 billion events remaining. 
This is because \name{fib} contains a recursive function that calls itself many times. 
Most of the events are \code{Call} events that get filtered out by the call stack optimization. 

\subsection{RQ4. Effectiveness of Replay Optimization}

\begin{table}[t]
\footnotesize
  \caption{Replay optimization experiment results.}
\label{t:replay}
\begin{tabular}{l|r|r|r|r||r|r|r|r}
\hline
\multirow{2}{*}{\textbf{Name}} &
  \multicolumn{4}{c||}{\textbf{Load+Validation time ($\mu$s)}} &
  \multicolumn{4}{c}{\textbf{Execution time ($\mu$s)}} \\ \cline{2-9}
 &
  \multicolumn{1}{c|}{\textbf{No}} &
  \multicolumn{1}{c|}{\textbf{Split}} &
  \multicolumn{1}{c|}{\textbf{Merge}} &
  \multicolumn{1}{c||}{\textbf{All}} &
  \multicolumn{1}{c|}{\textbf{No}} &
  \multicolumn{1}{c|}{\textbf{Split}} &
  \multicolumn{1}{c|}{\textbf{Merge}} &
  \multicolumn{1}{c}{\textbf{All}} \\ \hline\hline
\name{boa}             & 242742.4   & 102.87\% & 108.98\% & 106.05\% & 47262.4   & 101.87\% & 104.92\% & 101.62\% \\ \hline
\name{bullet}          & 27313.8    & 97.06\%  & 100.47\% & 104.62\% & 358420.8    & 101.55\% & 113.76\% & 127.99\% \\ \hline
\name{commanderkeen}   & 270972.2   & 116.73\% & 92.19\%  & 111.92\% & 18778942.4 & 100.96\% & 97.95\%  & 97.81\%  \\ \hline
\name{factorial}       & 2517.3     & 94.14\%  & 96.90\%  & 99.63\%  & 64.7      & 97.84\%  & 97.30\%  & 99.85\%  \\ \hline
\name{ffmpeg}          & 264556.5  & 100.81\% & 113.45\% & 120.35\% & 22.8        & 101.32\% & 106.80\% & 107.68\% \\ \hline
\name{fib}             & 5927.2    & 121.98\% & 118.52\% & 119.83\% & 58289332.4 & 99.99\%  & 99.13\%  & 98.15\%  \\ \hline
\name{figma-startpage} & 10758.8  & 98.93\%  & 101.38\% & 99.37\%  & 63.8        & 99.22\%  & 103.61\% & 98.75\%  \\ \hline
\name{game-of-life}    & 88.4 & 96.28\%  & 94.02\%  & 98.14\%  & 37.6 & 98.88\%  & 94.85\%  & 100.52\% \\ \hline
\name{guiicons}        & 17627.2   & 130.33\% & 113.69\% & 111.37\% & 522202.34   & 119.45\% & 120.20\% & 115.28\% \\ \hline
\name{hydro}           & 31932.9   & 97.84\%  & 99.85\%  & 98.65\%  & 84.6       & 89.24\%  & 94.32\%  & 90.54\%  \\ \hline
\name{jqkungfu}        & 29246.2   & 94.42\%  & 90.05\%  & 79.93\%  & 23.6       & 98.09\%  & 94.49\%  & 88.77\%  \\ \hline
\name{jsc}             & 218154.5   & 102.24\% & 84.84\%  & 87.97\%  & 19046.9     & 97.36\%  & 93.34\%  & 96.51\%  \\ \hline
\name{mandelbrot}      & 60570.8  & 131.97\% & 7.26\%   & 7.33\%   & 44074249.6 & 99.55\%  & 99.32\%  & 98.91\%  \\ \hline
\name{multiplyDouble}  & 6533.8    & 101.58\% & 106.38\% & 107.45\% & 43857353.6 & 100.05\% & 99.26\%  & 100.11\% \\ \hline
\name{multiplyInt}     & 5861.4     & 120.96\% & 113.67\% & 120.45\% & 42531426.8 & 100.22\% & 99.16\%  & 98.55\%  \\ \hline
\name{pacalc}          & 10414.5    & 90.26\%  & 92.22\%  & 103.58\% & 3453.8     & 92.95\%  & 93.85\%  & 102.23\% \\ \hline
\name{parquet}         & 78150.8   & 90.79\%  & 90.32\%  & 90.08\%  & 188.5     & 96.45\%  & 96.07\%  & 96.29\%  \\ \hline
\name{pathfinding}     & 21315.8    & 129.43\% & 123.20\% & 166.98\% & 3368370.8  & 101.60\% & 96.35\%  & 85.21\%  \\ \hline
\name{rfxgen}          & 27742.7    & 98.81\%  & 96.84\%  & 94.44\%  & 1146599.5   & 99.65\%  & 98.32\%  & 92.56\%  \\ \hline
\name{rguilayout}      & 26222.4   & 100.21\% & 98.27\%  & 97.52\%  & 440909.2   & 96.04\%  & 98.48\%  & 95.24\%  \\ \hline
\name{rguistyler}      & 21885.1    & 110.35\% & 114.29\% & 120.18\% & 469477.9    & 107.37\% & 120.39\% & 119.56\% \\ \hline
\name{riconpacker}     & 25251.2    & 95.74\%  & 84.12\%  & 87.78\%  & 16982.4    & 95.36\%  & 87.32\%  & 88.26\%  \\ \hline
\name{rtexpacker}      & 25133.0    & 87.92\%  & 79.43\%  & 78.43\%  & 15.2        & 86.18\%  & 86.18\%  & 83.22\%  \\ \hline
\name{rtexviewer}      & 16457.3    & 97.52\%  & 96.57\%  & 88.02\%  & 12.4        & 97.58\%  & 94.76\%  & 96.77\%  \\ \hline
\name{sandspiel}       & 22352.0      & 152.33\% & 141.32\% & 155.21\% & 3135717.0     & 105.89\% & 107.45\% & 92.61\%  \\ \hline
\name{sqlgui}          & 41837.3    & 103.06\% & 98.04\%  & 98.30\%  & 6790.0     & 100.99\% & 97.27\%  & 97.28\%  \\ \hline\hline
\textbf{Geomean} &
  \multicolumn{1}{l|}{} &
  \textbf{105.30\%} &
  \textbf{91.35\%} &
  \textbf{94.05\%} &
  \multicolumn{1}{l|}{} &
  \textbf{99.28\%} &
  \textbf{99.48\%} &
  \textbf{98.40\%} \\ \hline
\end{tabular}
\end{table}

In this section, we evaluate the effectiveness of replay optimization techniques introduced in Section \ref{s:replay-opt}.
We first conduct a simple comparison of the replay binary size of replay benchmarks 1) without any replay optimizations, 2) only with the memory write merge optimization, 3) only with the function split optimization, and 4) with both optimization techniques.
We found that the function split optimization does not affect the binary size, while the memory write merge optimization reduces the replay binary size by 9.98\%.
Then, for the 27 replay benchmarks in Wasm-R3-Bench, we carried out an ablation study to evaluate the effectiveness of the two optimizations.
We call this the \emph{replay optimization experiment}.
In the replay optimization experiment, we measure two kinds of times: \emph{load and validation time} and \emph{execution time}.
The load and validation time is the time spent loading, parsing, and validating the replay Wasm benchmark, while execution time represents time to execute the main (i.e. top-level replay) function.
We measure these with the \C{{-}{-}metrics} option of the Wizard engine, which reports load, validation, compilation, and execution time, and average over 10 runs.

Table~\ref{t:replay} reports the results of the replay optimization experiment.
On average, applying both replay optimizations reduces the time spent in load and validation by about 6\%.
In detail, applying the function split optimization increases the load and validation time by about 5\%.
Considering that some replay benchmarks require the function split optimization to run, we think this is an acceptable increase. 
Applying only the memory write merge optimization decreases the load time about 9\%.
Individually, \name{mandelbrot} seems to enjoy the greatest benefit, with the load and validation time reduced to just 7.33\% of the original time.
\name{rtexpacker}, \name{jqkungfu}, and \name{riconpacker} are other beneficiaries, with the load and validation time reduced to 78.43\%, 79.93\%, and 87.78\% of the original load and validation time, respectively.
On average, applying both replay optimizations decreases the execution time to 98.40\%.
We believe that although the replay optimizations do not significantly affect the execution time, it gives a slight performance improvement.
We are currently investigating the reasons behind.

Note that \name{funky-kart}, despite successfully completing the experiment, is excluded from the table.
This is because its replay benchmarks without replay optimizations exceeded the maximum size limit for function bodies imposed by the production Wasm engines, which prevented them from running on any engines.
Thus, we discuss the performance characteristics of \name{funky-kart} here.
While the function split optimization makes the replay benchmark executable, it does not affect its binary size.
Applying the memory write merge optimization on top of the function split optimization reduces the load and validate time to 7.80\% and the execution time to 86.31\% compared to the replay benchmark with only the function split optimization applied.
Thus, alongside \name{mandelbrot}, \name{funky-kart} is another example where the memory write merge optimization significantly reduces the load time and validation time of the replay benchmark, with an added benefit of reducing the execution time.

In summary, we show that the replay optimization techniques play a crucial role in reducing the load time and validation time of performance outliers in Wasm-R3-Bench, while having a modest effect on the execution time.

\section{Limitations}
\label{sec:limitations}

Currently, our Wasabi-based instrumentation supports Wasm version 2.0, with the exception of the SIMD proposal.
This limitation can be easily overcome by using a more up-to-date instrumentation library.
Among the proposals in phase 4, which are planned to be standardized soon, the multi-memory proposal is already supported by Wasm-R3.
We also believe our approach can support Wasm GC with some help from the host environment.
Our technique relies on making shadow copies of all mutable state and detecting modifications by comparing the shared state with the shadow state which necessitates host support, since Wasm \verb$funcref$ and \verb$externref$ do not have native Wasm comparison operators.
The most challenging proposal to support would be threads, which remains an open research question, as deterministic replay of racy programs lacks satisfactory and robust solutions.

\section{Related work}
\label{sec:relatetd}

Record and replay is a mature area of research that has been explored in various contexts, including architectural support for record and replay \cite{FDR}, operating system-level record and replay \cite{revirt,R2}, and language runtime-level record and replay~\cite{RANDR}.
Typically, such systems require intrusive modification to the respective CPU design, kernel, libraries, or language runtime to record events, but can also be done by bytecode rewriting~\cite{Versatile}.
In any case, potentially non-deterministic operations and side-effects must be identified and recorded as events.
Enumerating these operations can represent significant manual work.
Our approach has a different requirement in that we aim to run Wasm-R3 on any architecture, operating system, or language runtime without modification.
This requirement led us to using bytecode-level instrumentation.

Among previous instrumentation of this kind in the literature, Wasm-R3 is most similar to JSBench~\cite{jsbench}, a record and replay technique for the automated construction of JavaScript benchmarks.
The major difference between JSBench and Wasm-R3 stems from the design differences in the languages they target: JavaScript and Wasm, respectively.
In Wasm, instances and their host environment are cleanly separated by the import/export boundary.
Thus, efficiently tracking non-determinism caused by the host environment boils down to making shadow copies of the shared mutable states and comparing them.
In contrast, nearly any JavaScript operator could have unbounded side-effects.
For instance, JSBench notes that a JavaScript \verb|for...in| loop can be a potential source of non-determinism.
This leads the authors of JSBench to describe their catalogue of non-determinism in JavaScript applications as "necessarily incomplete."
We believe that the simplicity of Wasm-R3's record phase, in comparison to JSBench, is not a drawback but an advantage, demonstrating Wasm's suitability as a target for automatic benchmark generation for the web.

Jalangi~\cite{jalangi} is another record and replay framework for JavaScript, designed for heavy-weight dynamic analysis.
While Jalangi facilitates record and replay across different environments (e.g., recording on a mobile environment and replaying on a desktop environment), it operates within the confines of JavaScript engines, which accept the same inputs and produce identical outputs.
This means that, despite the convenience offered by Jalangi's record and replay functionality, it does not enable something fundamentally impossible; it is possible to have the same execution in different environments by simply running the same JavaScript code.
In contrast, Wasm-R3 enables the capability to replay a Wasm instance interacting with JavaScript code in non-web Wasm engines, which is impossible without Wasm-R3.
Moreover, the basic technique of Wasm-R3 could allow recording in non-web Wasm engines and replaying in web Wasm engines.
Another important difference is that, unlike Wasm-R3, Jalangi applies code instrumentation in both the record and replay phases to implement the replay of the recorded trace.
Although this eliminates the need to precisely determine at which point in the host code an interaction occurred, which Wasm-R3 meets, it has the disadvantage of mixing up the original code and replay code.
We instead aim to preserve the original binary and its exact instruction-by-instruction behavior and only replace calls to the host with replays.
Lastly, Jalangi explicitly states that they did not make any effort to optimize their implementation.
In contrast, we applied numerous optimizations to the trace to enable Wasm-R3 to scale to real-world applications, and to the replay IR, to ensure the resulting replay is representative of the recorded execution.

Other record and replay frameworks for JavaScript include Mugshot~\cite{Mugshot} and WebRR~\cite{WebRR}.
Mugshot~\cite{Mugshot} records browser events and shares our goal of recording on unmodified browsers.
Unlike Wasm-R3, Mugshot is tightly coupled with the web browser implementations and adopts different strategies depending on the browser.
In contrast, Wasm-R3 exploits the host-agnostic nature of Wasm to record the execution of Wasm applications across all browsers. 
WebRR~\cite{WebRR} proposes a record and replay technique that enhances the robustness of fragile end-to-end tests.
Their primary focus is on avoiding test failures that occur without a bug or misbehavior in the application under test.
In contrast, benchmarks generated from Wasm-R3 do not suffer from any kind of fragility that plagues end-to-end tests.
We refer to a comprehensive survey for a more detailed discussion of dynamic analysis for JavaScript~\cite{jsSurvey2017}. 

Beyond record and replay, a lot of research has gone into Wasm.
\citet{realWorldBinaries} proposed benchmarks of real-world binaries.
Unlike the benchmarks created with Wasm-R3, their benchmarks are not executable, making them unusable for performance benchmarking.
Several general-purpose techniques to dynamically analyze Wasm have been proposed, e.g., based on source-to-source instrumentation~\cite{Wasabi} or via dynamic instrumentation inside a Wasm engine~\cite{titzer2024flexible}.
Other work supports reverse engineering by inferring types~\cite{SnowWhite} and the purpose of functions~\cite{DBLP:conf/www/Romano023} in Wasm binaries,
studies security issues in Wasm~\cite{everythingOldIsNewAgain}, and
shows how to use Wasm for obfuscation~\cite{Wobfuscator}.
Our work contributes to the field by providing the first record and replay technique for Wasm and a benchmark suite of executable Wasm binaries.

\section{Conclusion}
\label{sec:conclusion}

We present Wasm-R3, the first record and replay framework for Wasm. The approach works with no modifications to the source compiler, virtual machine, host environment, operating system, or hardware.
During arbitrary execution of a Wasm module within a host environment, Wasm-R3 records all interactions with the host environment, detecting updates to shared mutable memory via a shadow memory.
The resulting trace is optimized to produce a precise replay binary that reproduces the original program's behavior by replaying all interactions with the host.
Wasm-R3 is broadly applicable to numerous real-world Wasm applications, and the replay files it generates can be run across both web and non-web Wasm engines.
Several optimizations applied at both recording and replay make infeasibly long traces tractable and reduce the overhead of trace replay for more faithful execution characteristics.
We have made Wasm-R3 available as open source and hope it will be beneficial to Wasm application developers to create repeatable replays of specific executions for benchmarking and other purposes.
In particular, we hope the technique can unlock a new era of Wasm benchmarking that better represents real-world use cases by routinely generating replays from real applications in their respective host environments.

\section*{Data-Availability Statement}
The artifact is available on Zenodo at \cite{baek_2024_13382344}.
It includes Wasm-R3 and the Wasm-R3-Bench benchmark suite.

\section*{Acknowledgments}
This work was supported by the European Research Council (ERC, grant agreement 851895), and by the German Research Foundation within the ConcSys, DeMoCo, and QPTest projects.
Also, this work was partly supported by
the National Research Foundation of Korea (NRF) (2022R1A2C2003660 and 2021R1A5A1021944) and
Institute for Information \& Communication Technology Planning \& Evaluation (IITP) funded by the Korean government MSIT (No. RS-2024-00337703).
Partial support also provided by the National Science Foundation under award \#2148301.

\bibliographystyle{ACM-Reference-Format}
\bibliography{references}

\end{document}